\documentclass[fleqn,usenatbib]{mnras}

\usepackage[T1]{fontenc}
\usepackage{ae,aecompl}

\usepackage{graphicx}
\usepackage{dcolumn}
\usepackage{bm}
\usepackage{hyperref}
\usepackage{graphicx}
\usepackage{fancyref}
\usepackage{amsmath}
\usepackage{subfigure}
\usepackage{tabularx}
\usepackage{threeparttable}
\usepackage{braket}
\usepackage{mathtools}
\usepackage{stmaryrd}
\usepackage{amssymb}
\usepackage{mathrsfs}
\usepackage{float}
\usepackage{natbib}

\title[Axion Structure Formation II]{Axion Structure Formation II: The Wrath of Collapse}

\author[E. W. Lentz et al.]{
Erik W. Lentz,$^{1,3}$\thanks{E-mail: erik.lentz@uni-goettingen.de}
Thomas R. Quinn,$^{2}$
Leslie J Rosenberg$^{3}$
\\
$^{1}$Institut f\"ur Astrophysik, Georg-August Universit\"at G\"ottingen, 
                 G\"ottingen, Deutschland 37707;\\
$^{2}$Department of Astronomy, University of Washington, 
                 Seattle, WA, USA 98195-1580;\\
$^{3}$Department of Physics, University of Washington,
                 Seattle, WA, USA 98195-1580;
}

\date{Accepted XXX. Received YYY; in original form ZZZ}

\pubyear{2018}

\begin{document}
\label{firstpage}
\pagerange{\pageref{firstpage}--\pageref{lastpage}}
\maketitle

\date{\today}

\begin{abstract}

The first paper in this series showed that QCD axion dark matter, as a highly-correlated Bose fluid, contains extra-classical physics on cosmological scales. The nature of the derived extra-classical physics is exchange-correlation interactions induced by the constraints of symmetric particle exchange and inter-axion correlations from self-gravitation. The paper also showed that the impact of extra-classical physics on early structure formation is marginal, as the exchange-correlation interaction is inherently non-linear. This paper continues the study of axion structure formation into the non-linear regime by now considering the case of full collapse and virialization. The N-body method is chosen to solve the collapse problem, and its algorithms are derived for a condensed Bose fluid. Simulations of isolated gravitational collapse are performed for both Bose and cold dark matter fluids using a prototype N-body code. Unique Bose structures are found to survive even the most violent collapses. Bose post-collapse features include dynamical changes to global structures, creation of new broad substructures, violations of classical binding energy conditions, and new fine structures. Effective models of the novel structures are constructed and possibilities for their discovery are discussed.

\end{abstract}



\section{Introduction}
\label{Introduction}

The first paper of this series (\citet{Lentz2018b}, ASF1) presents a new model of structure formation for relic QCD axions that more fully details the correlated nature of condensed self-gravitating Bose fluids in a FLRW (Friedmann-Lema\^itre-Robertson-Walker) cosmology \citep{MTW}. Such a model did not exist in the literature previously, due to contention over the appropriate scope of condensation for a gravitationally interacting medium. The various perspectives in this debate lead to drastically different conclusions on the existence of axion structures distinct from collision-less cold dark matter (CDM), with many supporting the position that unique structure does not exist beyond the de Broglie scale (for instance, see \citet{Berges2015,Davidson2015,Guth2015}), and others promoting the existence of super-de Broglie effects \citep{Sikivie2009,Erken2012,Banik2016}. The Axion Structure Formation series of papers aims to resolve the extent to which degeneracy, correlation, and quantum mechanics beyond the mean field can create structure in self-gravitating bosonic dark matter that are unique from CDM.

\begin{figure}
\begin{center}
\includegraphics[width=9cm]{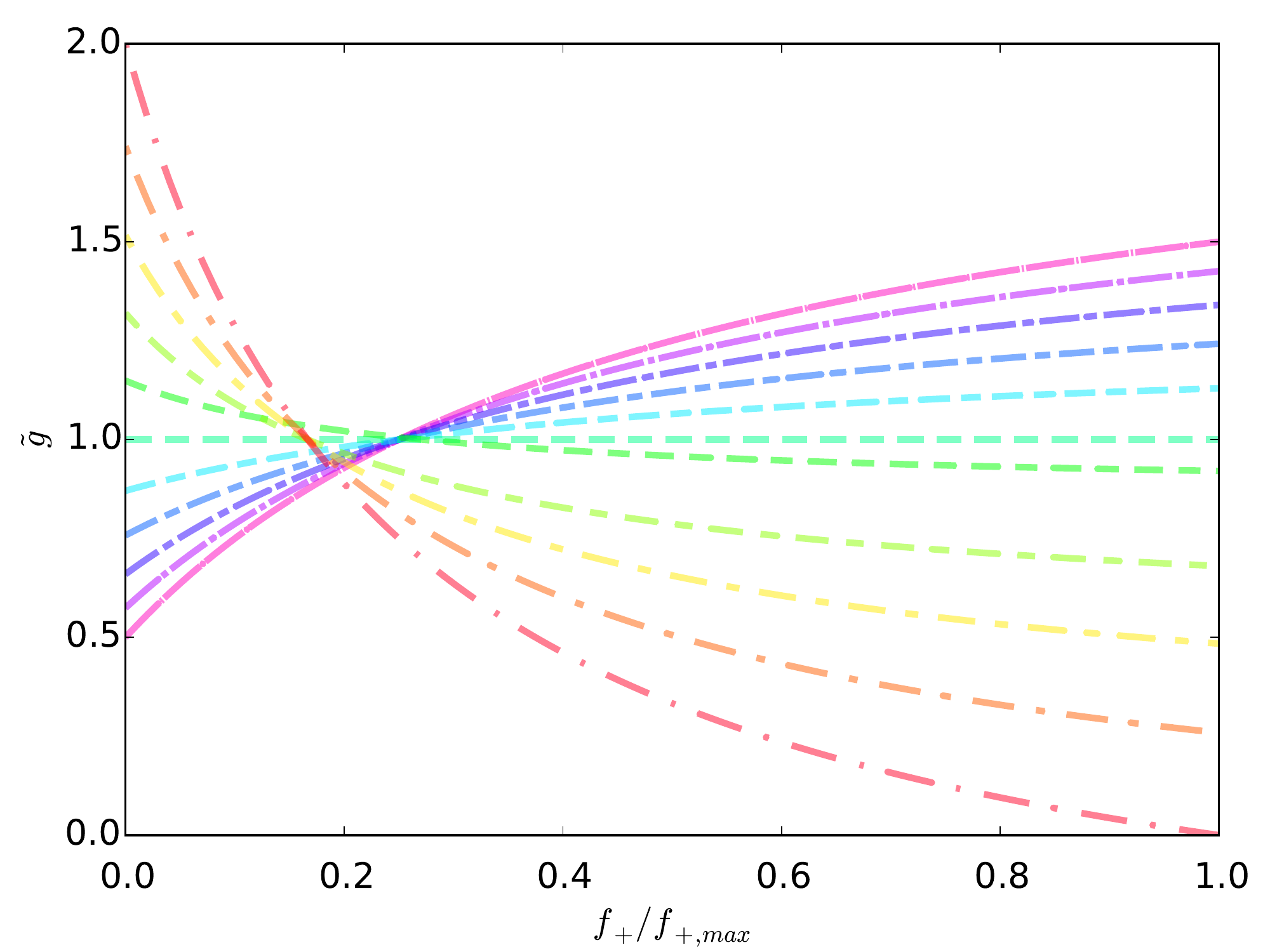}
\caption{Correlation functions for several types of systems, both bosonic and fermionic, taken from Figure C1 of \citet{Lentz2018b}. The lines that lie below unity at the maximum of the distribution function $f_+/f_{+,max}=1$ (red to green; dash-dotted to dashed) are from non-interacting Fermi distributions, organized in a spin-symmetric configuration, parameterized by temperature decreasing with initial correlation like $T = T_f/(C-1)^2$, where $T_f$ is the Fermi temperature, and $C$ is the correlation at zero density. The lines at and above unity (green to violet; dashed to near-solid) are from condensed scalar Bose distributions shaped into a spherical Gaussian in phase space, and also become colder with more extreme initial correlation via the dispersion length relation $\tilde{\sigma}_v/\tilde{\sigma}_x \propto C$, where $\tilde{\sigma}_x$ and $\tilde{\sigma}_v$ are the position and velocity dispersions. $f_+$ is formally defined in Eqn.~\ref{fplus}.}
\label{gplot}
\end{center}
\end{figure}

ASF1 and \citet{Lentz2018a} (LQR) begin the process by providing a realistic model of weak-gravity non-relativistic condensate dynamics, mapping the covariant quantum theory of the axion field into many-body quantum mechanics after the radiation-dominated era. By constructing a Hamiltonian of the co-moving non-relativistic many-body axion system, explicit solutions of the many-body state are found in ASF1 in the Compton-Newtonian limit. A more precise definition of condensation can also be formed in this context, which differs subtly but significantly from standard mean field theory (MFT). The resulting motion of such a maximally-condensed fluid is found to be different from a pressure-less fluid, MFT, or classical field, even on super-de Broglie length scales, owing to its conformal two-body correlation function. Also, it is worth noting that self-gravitating axion MFTs and classical fields in the literature produce pressure-less fluids in the super-de Broglie limit \citep{Berges2015,Davidson2015,Guth2015,Mocz2018,Veltmaat2018}. Sample behavior of the correlation function from ASF1, for simple Bose condensates and Fermi fluids near the condensed limit, can be viewed in Fig.~\ref{gplot}. As there are no intrinsic scales for this purely self-gravitating system, the departure from unity of the dimensionless correlation function is measured by global deviations from single particle separability. 

The model laid out in ASF1 contains new physics and the potential for new dark matter structures inherited from a highly-correlated axion condensate, on conceivably any scale so long as the Newtonian and super-de Broglie conditions hold. The self-gravitating exchange-correlation interactions are inherently non-linear, making them invisible in linear analyses of an early universe close to the homogeneous and separable limits. This null result is demonstrated in ASF1 using linear perturbation theory on an otherwise standard depiction of structure near the era of recombination. Additional allowances are required to be given to the extra-classical forces in order to separate the correlation effects formally from zero.

This paper addresses whether or not the unique physics in Bose condensate dynamics can impact the more violent phases of structure formation. Namely, do unique Bose structures form during non-linear gravitational collapse, through shell crossing, and survive into virialization. Analytical techniques cannot be relied on for such computations, but an appropriate numerical method can accurately and stably perform these calculations. The N-body method, and more generally the method of characteristics, will be the numerical technique of focus for this paper, due to its success in solving for classical infall \citep{Courant1953,Dehnen2011}. N-Body methods have proven quite useful in calculating all phases of dark matter (DM) structure formation to date \citep{Diemand2005,Angulo2012,Pillepich2018}. N-Body methods are particularly well-suited for modeling CDM, as they can capture the diffuse dynamics and reliably model force fields of collision-less fluids in phase space over many orders of magnitude. Further, many sophisticated and highly-scalable platforms already exist to simulate such systems. It should be straight-forward to incorporate Bose infall dynamics into such a platform.

It is prudent to first test the Bose model in a simple implementation prior to sinking significant resources into the development, testing, and detailed simulation using a large and complex platform. This second paper of the axion structure formation series derives an N-body method for Bose infall and tests it in a simple implementation on a range of collapses using a prototype code. The results of these simulations give us a preliminary indication of whether or not structures unique to Bose fluids can survive gravitational collapse. The remainder of this paper is structured as follows: Section~\ref{CSF} reviews the continuum model of axion structure formation; Section~\ref{NBA} converts the continuum description into an N-Body algorithm; Section~\ref{CXP} introduces the small-scale N-Body code, its gravity solver, the suite of initial conditions, and the series of preliminary simulations performed for both classical and Bose gravitational collapse; Section~\ref{Results} presents the results from the simulation series, showing that unique structures not only exist in, but in some cases thrive on, violent collapse; Section~\ref{Discussion} discusses the simulations and begins to build some intuition for the novel Bose structures found, exploring several signals for search; and Section~\ref{Summary} presents prospects for further investigation, including the incorporation of these algorithms into one of the large-scale N-Body codes.

\section{Condensate Structure Formation}
\label{CSF}

The many-body quantum state of $N$ relic axions past the matter-radiation transition is governed by a co-moving Schr\"odinger equation of dimension $3N + 1$. ASF1 and LQR found solutions of this equation to be more compactly described by inter-axion correlators than the standard single product Fock space representation used by many MFTs. Even with this insight, a general galactic halo contains far too many axions to simulate without a reduction in the degrees of freedom. ASF1 and LQR also found it possible to reduce the dynamics of the full system to tracking the motions of a single-body density. This reduction is performed in ASF1 and LQR using a Wigner transformation \citep{Wigner1932}, converting the wave-function representation to a pseudo-distribution function (pDF) over phase spaces, then integrating over $N-1$ phase spaces. The impact of  position-momentum complementarity is trivialized by the super-de Broglie limit in which the halo is evaluated, effectively converting the Wigner function to a true DF. Fortunately, modern cosmological simulations are smoothed over scales on the order of many parsecs, far above the expected QCD axion wavelengths, removing the influence of de Broglie-level dynamics. The only component of the correlated self-gravitating many-body state to survive super-de Broglie smoothing and integration over all but a single phase space is the conformal two-body correlation function
\begin{equation}
\tilde{g} = \frac{C+\lambda_1 f_+}{1+ \lambda_2 f_+},
\end{equation}
where $C$ is a measure of the axion's correlation at zero density, $\lambda_1$ and $\lambda_2$ are Lagrange multipliers inherited from the correlation function construction  of ASF1 and LQR, and $f_+$ is a two-body function composed of the symmetric combination of single-body DFs
\begin{equation}
f_+ = \frac{1}{2}\left(f^{(1)}(w_1,t) + f^{(1)}(w_2,t) \right). \label{fplus}
\end{equation}

The dynamics of a completely condensed axion DF are governed by an equation of motion of Boltzmann-like form, and a Poisson equation for the mean field gravitational potential. Evolving in the presence of a FLRW cosmology in the large $N$ limit, the set of equations in co-moving coordinates becomes
\begin{align}
&0 = \partial_t f^{(1)} + \frac{\bmath{v}}{a^2} \cdot \bmath{\nabla} f^{(1)} -  \bmath{\nabla} \Phi' \cdot \bmath{\nabla}_v f^{(1)} -  \bmath{\nabla} \bar{\Phi} \cdot \bmath{\nabla}_v f^{(1)} \nonumber \\
&- m_a \bmath{\nabla}_v \cdot \Bigg(   \int d^6 w_2 f^{(1)}(w_2,t)  \bmath{\nabla} \phi_{12} \nonumber \\
& \times\left( \frac{C-1 - \left(\lambda_1+\lambda_2\right)f_+}{1+ \lambda_2 f_+} \right) f^{(1)}(w_2,t)\Bigg), \label{Boltz} \\
& \nabla^2 \bar{\Phi} = 4 \pi G Ma^2 \int d^3v f^{(1)}, \label{Poisson}
\end{align}
where $a$ is the FLRW cosmology scale factor, $G$ is the Newtonian gravitational constant, $m_a$ is the mass of an axion, $M$ is the total system mass,  $f^{(1)}$ is the single-body DF, $\Phi'$ is the Newtonian gravitational potential from non-axion or non-condensed species. The non-axion species have been added to the Boltzmann-like equation in the canonical way. The gravitational potential of the non-axion species in general is governed by their own Poisson equation. $\bar{\Phi}$ is the single-body averaged Newtonian gravitational potential from axions, and $\phi_{12}$ is the inter-axion gravitational kernel
\begin{equation}
\phi_{12} = \phi(\bmath{x}_1,\bmath{x}_2) = - \frac{G a^2 }{ |\bmath{x}_1-\bmath{x}_2|}.
\end{equation}

 The Lagrange multipliers $\lambda_1$ and $\lambda_2$ of the inter-axion correlation function are set by the normalization constraints on the single-body DF and the correlation function. The constraints are encapsulated by the single expression
\begin{equation}
1 = \int d^6 w_1f^{(1)}(w_1,t) \frac{C-\lambda_1 f_+}{1+ \lambda_2 f_+}, \label{lc}
\end{equation}
which is a function over phase space $w_2$. Lastly, it is often convenient in the equation of motion to replace the multiplier $\lambda_1$ with $\lambda_+ = \lambda_1 + \lambda_2$. The $(\lambda_+,\lambda_2)$ pair will be the normal convention for the remaining sections. We refer to the Bose-specific dynamical contribution as exchange-correlation (XC) dynamics since they exist due to the enforcement of the exact Bose particle exchange symmetry and inter-axion correlation from self-gravity. The collection of Eqns.~\ref{Boltz}, \ref{Poisson}, \ref{lc} is the starting point for constructing the axion N-Body algorithm.

\section{N-Body Axions}
\label{NBA}

An efficient numerical algorithm breaks down all continuum features of the axion model into discrete, well-ordered algebraic steps. The algorithm chosen for this paper is a combination of the method of characteristics (MOC) and leapfrog integration. MOC is a technique of discretizing over the space-like dimensions of the hyperbolic Boltzmann-like equation orthogonal to the time direction, here represented by phase space, by which we parametrize the characteristic curves of the solving DF \citep{Courant1953}. The symplectic leapfrog method is used to integrate over the remaining time-like direction in a way that is stable to integrals of the motion \citep{Ruth1983}. Derivations are restricted to purely condensed axion DM, free from state diffusion. A detailed derivation of the algorithm may be found in Appx. \ref{MOC}. We assume here no interactions outside of self-gravity.

The algorithm first breaks down the spatial dimensions into a set of dynamic Lagrangian sample points, evolving through time via Hamilton-like equations, each of the form
\begin{align} 
\dot{f}_1 &= 0, \\
\dot{\bmath{x}} &= \frac{\bmath{v}}{a^2 }, \\
\dot{\bmath{v}} &= - \bmath{\nabla} \bar{\Phi} \nonumber \\
& - m_a \frac{\partial}{\partial \bmath{\nabla}_{v} f_1} \int d^6w_2 \bmath{\nabla} \phi_{12} \cdot \bmath{\nabla}_{v} \left(f_1 \frac{C-1 - \lambda_+ f_+}{1 + \lambda_2 f_+} f_2\right) ,
\end{align}
where $f_i$ is the single body DF over the i-th phase space, $w_1 = (\bmath{x},\bmath{v})$, and $\partial/(\partial \bmath{\nabla}_{v} f_1)$ is a functional derivative. Each sample can naturally be represented by the mass fraction $m$ enclosed within its representative volume in phase space. For clarity, the functional derivative evaluates to
\begin{align}
    &\frac{\partial}{\partial \bmath{\nabla}_{v} f_1} \int d^6w_2 \bmath{\nabla} \phi_{12} \cdot \bmath{\nabla}_{v} \left(f_1 \frac{C-1 - \lambda_+ f_+}{1 + \lambda_2 f_+} f_2\right) = \nonumber \\
    & \int d^6w_2 \bmath{\nabla} \phi_{12}  \left( \frac{C-1 - \lambda_+ f_+}{1 + \lambda_2 f_+} f_2\right) \nonumber \\
    &+ \int d^6w_2 \bmath{\nabla} \phi_{12}  \left(f_1 \frac{- \lambda_+/2}{1 + \lambda_2 f_+} f_2\right) \nonumber \\
    &- \int d^6w_2 \bmath{\nabla} \phi_{12}  \left(f_1 \frac{\lambda_2/2 (C-1 - \lambda_+ f_+)}{(1 + \lambda_2 f_+)^2} f_2\right).
\end{align}
The resemblance of the sample equations to classical particle motion is striking, with the extra-classical deviations looking like additional forces acting upon the Bose `particles'. Sample particles' equations of motion are accurate to the precision of the forces. 

The second step of the algorithm integrates the sample points over time by breaking down the continuous progression into an algebraic sequence on discrete time steps. Over a single period of time $T = [0,t]$, the leapfrog method takes three operations, often referred to as drift-kick-drift, to calculate a sample's next phase space configuration
\begin{align}
\bmath{x}(t/2) &= \bmath{x}(0) + \frac{\bmath{v}(0)}{a^2 } \frac{t}{2} ,\\
\bmath{v}(t) &= \bmath{v}(0) + t \Bigg(- \bmath{\nabla} \bar{\Phi}  \nonumber \\
 - m_a & \frac{\partial}{\partial \bmath{\nabla}_{v} f} \int d^6w_2 \bmath{\nabla} \phi_{12} \cdot \bmath{\nabla}_{v} \left(f_1 \frac{C-1 - \lambda_+ f_+}{1 + \lambda_2 f_+} f_2\right) \Bigg) \Bigg|_{\bmath{x}(t/2)} ,\\ 
\bmath{x}(t) &= \bmath{x}(t/2) + \frac{\bmath{v}(t)}{a^2} \frac{t}{2},
\end{align}
where the shortened functional derivative form is kept for convenience. It is fortunate that the canonical form of leapfrog can be used despite the general velocity dependence of the XC force. The integration scheme is second-order in time accuracy, however the algorithm's symplectic nature does improve the long term accuracy and stability of quantities such as energy and angular momentum, even relative to many higher-order methods.

Calculation of the self-gravity can be performed in several ways. Once an optimized sampling choice has been made, so as to source the forces, an interpolation scheme of those forces can be constructed. Many efficient cosmology codes reduce the computational cost of gravity by differentiating between direct and long-range calculations of the classical gravitational potential, and simplifying the long-range calculations. This decomposition is often implemented on either a background mesh or a tree decomposition of particles. These interpolating techniques are capable of reducing the $O(n^2)$ scaling of direct two-body force calculation to a far more manageable $O(n \log n)$, or possibly in some cases even to $O(n)$ \citep{Dehnen2014}. Exchange-correlation interactions extend the force calculation in Bose fluids. Fortunately, the XC interactions can be interpolated similarly to Newtonian gravity in several common techniques, such as a tree decomposition, and preserve the scaling with particle number. Additional cost of the XC calculations are initially estimated at $\sim 50\%$ of Newtonian gravity. Other methods for gravity calculation also exist, some of which may be better able to resolve the fine structure of DM halos \citep{Hahn2016,Sousbie2016,colombi_alard_2017}. It remains to be seen whether XC forces can be integrated into these other methods.

\section{The Condensate in eXternal Potential Code and Simulations}
\label{CXP}

Developing a small-scale simulation code is important for testing the new infall model. Fast turn-around allows for a quick development cycle. Further, a small platform is better able to keep pace with the developing theoretical science, discussed further in Section~\ref{Summary}. Prior to implementing the above algorithms in one of the existing high-performance codes, we test them in a prototype platform, Condensate in eXternal Potential (CXP), to simulate isolated 3+1-dimensional systems of interacting degenerate bosons in isolation, over a static cosmological background. The resultant algorithms for XC force integration are designed such that they, in principle, hold for a more complex implementation when calculated in parallel with the Newtonian gravity counterpart. Here, CXP is used to simulate the gravitational infall of  near-spherical axion distributions through collapse and into virialization. A detailed layout of CXP is deferred to a forthcoming paper, though the algorithms relevant to this study are spelled out below.

\subsection{Parameters and Algorithms}
\label{params} 

CXP utilizes the N-body and time-integration algorithms of Section \ref{NBA} to simulate the phase-space evolution of self-gravitating bosons. The code is supplied an initial distribution of sample points and the initial correlation $C$ parameter. Sample points are quasi-randomly generated using a glass seed distribution \citep{Baertschiger2002}, which is mirrored, randomly sampled, and re-scaled to fit the simulation parameters of particle number and distribution shape.  Masses of each sample point and the initial correlation are propagated as integrals of the motion. Leapfrog evolution iterates over the Cauchy data, evolving it in time for a specified period. 

The value of the XC Lagrange multiplier parameters need only be calculated at the initial distribution as the $\lambda$s are also integrals of the motion. The constraint condition of Eqn.~\ref{lc} can be approximated via a mean-point integration
\begin{equation}
    0 = \sum_i^n w_i \frac{C-1 - \lambda_+ (f_i + f_2)/2}{1 + \lambda_2 (f_i + f_2)/2}, \label{lambdaalg}
\end{equation}
where we have canceled out the single distribution function integral to make the constraint condition null, and the $w_i$ are partitions of the phase space measure, or weights associated with each sample point. The weights play a similar role to the sample mass. Sample weights are taken to be equal across the distribution $w_i = w$, as are the masses of each particle. One can find the unique solution to both multipliers by evaluating Eqn.~\ref{lambdaalg} at two unique points of the single body distribution, $f_2$. CXP uses $f_2$'s maximum and minimum values.

The CXP implementation of gravity has not yet been explicitly stated. The primary advantage of the highly-parallelized routines, such as a tree decomposition, appear when used on large communications-limited networks, and are not of the same advantage for small $n$ systems. CXP uses the direct $O(n^2)$ approach to interaction calculation, in the interest of developmental efficiency and agility. 

For interaction evaluation, using the Coulomb kernel for a particle's gravitational potential is not appropriate, as each `particle' represents many axion quanta. Instead, a softened potential using the K1 profile of \citet{Dehnen2001} is issued for each sample. K1 has the effective Poisson-derived density
\begin{align}
&\Phi^{K1}(\acute{r}) =  \nonumber \\
&\Bigg\{
	\begin{array}{ll}
		\frac{-G}{32 \acute{r}} \left(64 - 105 \acute{r} + 175 \acute{r}^3 - 147 \acute{r}^5 + 45 \acute{r}^7 \right)   &  0 \leq r' \leq 1, \\
		-G/\acute{r}  & r'\geq 1,
	\end{array}
 \label{K1}
\end{align}
where $\acute{r}=r/d$, $d$ is the characteristic softening length, and the kernel mass has been set to unity. Note that for Bose systems simulated beyond the de Broglie scale, this softening should match the expected level of smoothing sufficient for minimal force error and not some intrinsic particle scale \citep{Dehnen2001}. The K1 kernel is chosen for its computational efficiency and compact support. The K1 kernel does have the odd characteristic of containing a region of negative Poisson mass density. This trait does not imply that the DF becomes negative in an annulus about the sample, it is merely an alteration to the force profile. A softening length commensurate with the particle density is used 
\begin{equation} 
d =  \epsilon \left( \frac{4/3 \pi R^3}{N} \right)^{1/3}, \label{softlength}
\end{equation}
where $R$ is the typical radial length scale of the initial configuration, and $\epsilon$ is a numerical factor. The K1 profile is not ubiquitous among pure N-Body codes, with some instead favoring a spline mass profile \citep{Merritt1996,Dehnen2001}.

The XC force is straightforward to calculate once the integral over
phase space is performed (see Appx.~\ref{MOC}).  Furthermore, the XC forces between
two particles are central, and only dependent on the particles'
positions, masses, and sample DF values, so that both angular momentum and energy are conserved in a collision-less Bose system. The total force felt by an N-body particle is
\begin{align}
    &\bmath{F}_1 = -\sum_i^n \bmath{\nabla} \Phi_{1i}\nonumber \\
    &- \sum_i^n \bmath{\nabla} \Phi_{1i}  \left[ \frac{C-1 - \lambda_+ \left( f_1 + f_i\right)/2}{1 + \lambda_2  \left( f_1 + f_i\right)/2}\right] \nonumber \\
    &- \sum_i^n \bmath{\nabla} \Phi_{1i}  \left[ \frac{- \lambda_+ f_1 /2}{1 + \lambda_2  \left( f_1 + f_i\right)/2} \right] \nonumber \\
    &+ \sum_i^n \bmath{\nabla} \Phi_{1i}  \left[ f_1 \frac{\lambda_2/2 \left(C-1 - \lambda_+  \left( f_1 + f_i\right)/2)\right)}{\left(1 + \lambda_2  \left( f_1 + f_i\right)/2 \right)^2} \right], \label{CXPforce}
\end{align}
where $\Phi_{1i}$ is the softened potential between samples at $\bmath{x}_1$ and $\bmath{x}_i$. CXP uses the same softening for XC as the mean Newtonian interaction. The first sum of Eqn.~\ref{CXPforce} gives the mean Newtonian force, and the last three sums are XC components. Recall that the sample DF value is integral along the characteristic curve, making the bracketed terms easily calculable when the initial DF value is propagated with the sample. This ease in computation obfuscates a subtle point, however, that the XC force can be velocity dependent. The terms in square brackets will be constant for a given particle pair because the phase space density is an integral of the motion in a purely condensed collision-less system.  However, collisions or a change in the system's density of states would change the particles' phase space densities over time.  In particular, collisions that change the velocity distribution will change the phase space density and hence introduce an implicit velocity dependence into the XC force on each particle. Mean Newtonian gravity would not respond to such changes. Our algorithm does not currently handle a dynamical density of states nor the collisional situation. Further discussion of the phase-space dependence of the XC force can be found in Appxs.~\ref{MOC}, \ref{SShell}.

Time-stepping is performed uniformly over the system for simplicity. The leapfrog time-step size is chosen such that the fastest orbits would contain multiple dozen steps so as to be well resolved at first shell crossing, even in the most violent of collapses. Such a requirement brings the time-step size to $\delta t \lesssim 10^{-2} t_{dyn}$ for many of the presented simulations, where $t_{dyn} = 1/\sqrt{G M/R^3}$ is the dynamical crossing time for a typical orbit of the uncorrelated system. The fast time step ensures adequate resolution of the fastest orbits in the system, but over-resolves the less dynamical regions. The code is currently parallelized over interaction calculation at the single time-step level over both single-node CPU and GPU resources.

We choose the duration of the simulations to be long enough that
virialization is well established for much of the bound portions of the halo, but short compared to the two-body relaxation timescale. The two-body relaxation timescale is
\begin{equation}
    t_{relax} \approx \frac{0.1 n}{ln(n)}t_{cross},
\end{equation}
as derived in \citet{BT2008}, where $t_{cross}$ is the system crossing time. The crossing time, in our case, is altered by the presence of additional forces, as is the form of the net force. We find the time to two-body relaxation, using an adjusted relaxation argument to account for XC, to be on the order of many dozens to thousands of crossing times for systems of $n=10^3-10^5$. Classical infall is naturally found to take the longest for fixed $n$ as the XC force on average amplifies the experienced gravitational force, shortening the crossing time by an estimated $1.5$ times at $C=0.5$. Only several dynamical times are required for a system to virialize, so a period of ten dynamical times is chosen to preserve the details of infall.

\subsection{Spherical Collapse Simulations}
\label{SCS}

The collapse of spherical or near-spherical distributions is a productive first step in classifying the effects of self-gravity. Several configurations are tested here, including the cold homogeneous sphere, the normal cold sphere, and versions imparted with small solid-body spin. Simulations consist of $50,000$ particles each. The configurations sampled for the suite of simulations presented here take on the outer product of parameters of shape, correlation, and spin
\begin{enumerate}
\item shape $\in \{$ Top-hat, Gaussian $\}$
\item $C \in \{0.5, 0.75, 0.9, 0.95, 1.0 \}$
\item $\lambda \in \{0.0,0.05,0.10\}$
\end{enumerate}
Individual simulations of this type will be referred to as $S(\text{shape},C,\lambda)$. CDM and Bose runs share initial DFs.

\subsubsection{Top-Hat Sphere}

The most basic and most violent collapses available to pressure-less self-gravitating fluids occur with sphere initial conditions of homogeneous density and zero velocity \citep{Gunn1972}. Density of the sphere is given by the normalized Top-hat distribution
\begin{equation}
\rho_{TH} = \frac{3 M}{4 \pi R^3}H(R-r),
\end{equation}
where $R$ is the radius of the sphere surface, $M$ is the total mass, and $H()$ is the Heaviside function. Giving the sphere a perfectly cold momentum distribution is tempting, but is to be avoided as it leads to divergent values of the Lagrange multipliers and also to numerical artifacts. A normally-distributed smoothing in momentum is applied, with momentum dispersion $\sigma_v$ being a small fraction of the crossing time. In the pressure-less fluid case, the cold homogeneous sphere ideally collapses to a single point before the drastic non-linearity of the singularity scatters the symmetry-breaking sampled distribution into a gentler configuration. Observing such severe collapses should provide a near-optimal probe of the XC contributions to the $1/R^3 \sigma_v^3$ or $1/d^3 v_{sng}^3$ scales. The time-step size is set at $\delta t = 0.5 \times 10^{-2} t_{dyn}$ so as to adequately resolve the softened singularity. A softening length factor of $\epsilon=2$ is used.

\subsubsection{Gaussian Sphere}

A less violent and more cosmologically representative example is the Gaussian sphere. Its density is given by an isotropic normal distribution
\begin{equation}
\rho_{G} = \frac{M R^3}{(2 \pi)^3} e^{-r^2/( R^2/2)},
\end{equation}
where $R/2$ gives the dispersion radius of the sphere. 

There exist instabilities for any simulations of cold near-spherical distributions occurring in three spatial dimensions. The well-studied radial orbit instability (ROI) \cite{Barnes1986,Valluri2000,Bellovary2008,Lentz2016} is known to produce a prolate bar feature through the distribution center. Bose physics may leave a signature in this symmetry-breaking structure, both in the homogeneous and normal cases. The normal sphere shares its other qualities such as primordial velocity dispersion with the homogeneous sphere. A time-step of $\delta t = 0.5 \times 10^{-2} t_{dyn}$ and a softening length factor of $\epsilon=1$ are used.

\subsubsection{Spinning Spheres}

Angular momentum plays a significant role in the formation of structure in cosmology, and understanding the reaction of the XC physics to angular stimuli is important for building an intuitive picture of Bose infall. It is not expected to see the influence of vortices, standard structures in rotating super-fluids, as they are suppressed here by a factor of the axion de Broglie scale. Angular momentum, instead, is to be imparted to the particles as a solid-body rotational boost about the center of the initial sphere
\begin{equation}
\bmath{v}_{boost} = \omega \hat{\bmath{z}} \times \bmath{r},
\end{equation}
where $\omega$ is the angular speed of the boost, $\hat{\bmath{z}}$ is an arbitrarily-chosen constant unit vector field of Euclidean space, and the multiplication operator is the canonical three-dimensional cross product. The self-gravitating spin parameter of \citet{Peebles1969} is used to calculated $\omega$ by gauging the relative size of angular momentum to the overall energy contribution
\begin{equation}
\lambda = \frac{J |E|^{1/2}}{G M^{5/2}},
\end{equation}
where $J$ is the magnitude of the total angular momentum, $E$ is the total binding energy, and $M$ is the total mass of the distribution. The classical CDM case retains an ROI-like feature for cosmologically-expected values of the spin parameter $\lambda$ \citep{Lentz2016}. Studying the outer product of simulation shape, correlation, and spin begins to fill the dictionary of potential unique Bose structures.

\section{Results}
\label{Results}

The collection of spherical collapse simulations show structural differences between Bose and classical infall along multiple dimensions. The observed Bose features broadly fall into three categories: shifts in global structure, creation of new broad substructures including violations of classical binding energy conditions, and new fine structure. Projections of the simulation output shown here may contain more than one of these features. Convergence tests for a sample of structures can be found in Appx.~\ref{ConvStdy}.

\subsection{Global Features}
\label{GF}

\begin{figure*}
\begin{center}
\includegraphics[width=\textwidth]{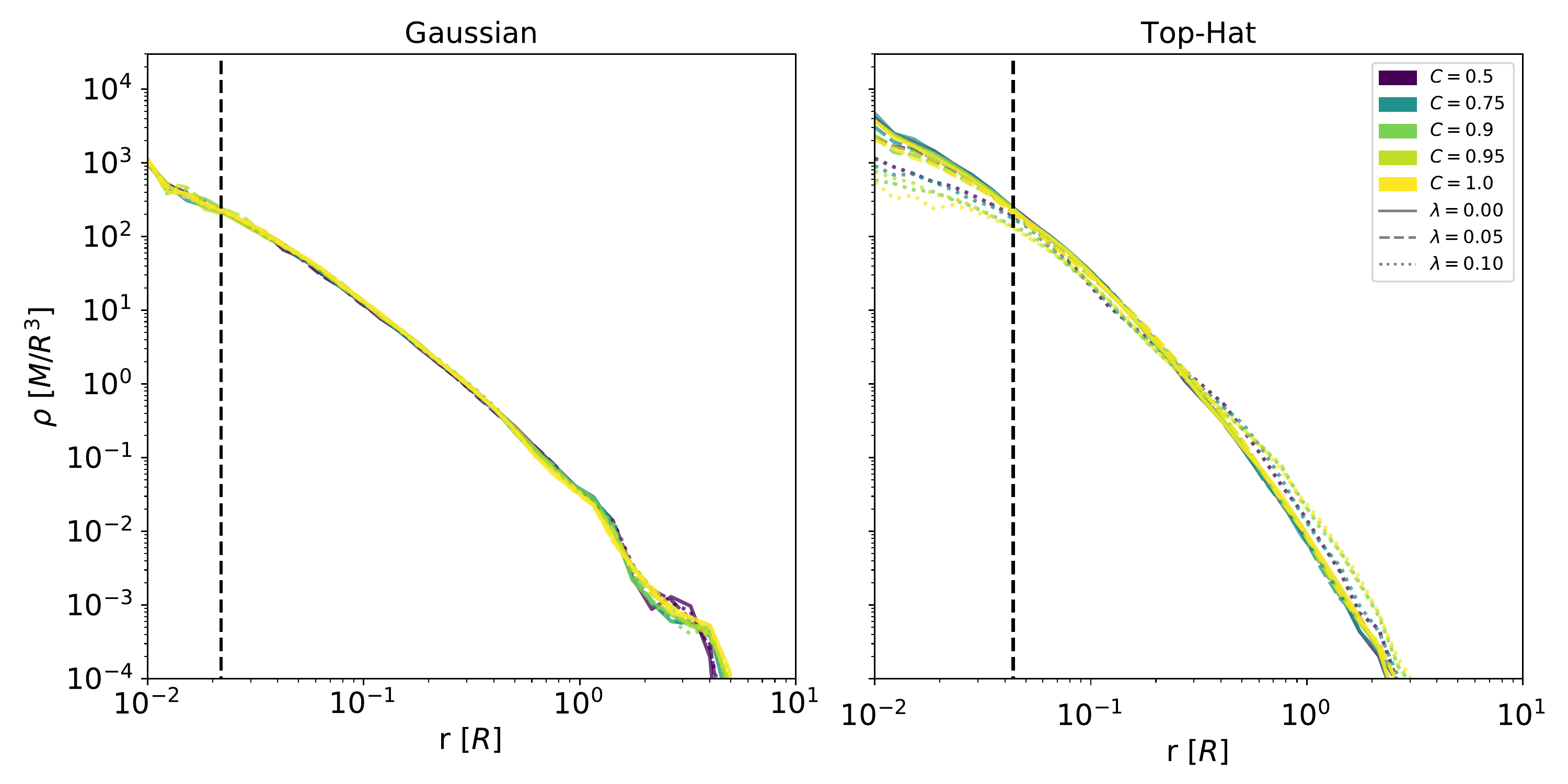}
\caption{Mass density radial profiles of the isolated collapse simulations of Sec.~\ref{SCS}. Gaussian profiles on L, Top-hat profiles on R. Halos were measured after evolving for $10 t_{dyn}$. Profile coloration indicates degree of correlation ranging from classical $C=1.0$ to highly correlated $C=0.5$. Line style indicates level of Peeble's spin $\lambda$ of the halo. The softening profile's maximum force radius is represented by the black dashed line in each simulation set, below which our confidence in the results are diminished.}
\label{fig:rhor}
\end{center}
\end{figure*}

The standard global measure of halo structure displays little susceptibility to XC. Radial density profiles of Gaussian infall show a universal broken power law behavior similar to the cosmologically-universal shape of NFW (Navarro-Frenk-White) \citep{Navarro1996a,Navarro1996b}, Fig.~\ref{fig:rhor}. The halos with static background show a steeper central cusp of $\alpha \approx -1.8$, breaking slowly at $r_{break} \sim 0.4 R$ into an outer power law of $\beta \sim -2.7$. Some small changes organized by correlation are seen at and beyond the virial radius, which occurs at $\sim 2-3 R$. The response of the new scale to spin is marginal, much like in cosmological CDM. No significant new structure in the form of a central core or other features are seen for Bose Gaussian collapse. This invariance is telling of the profile's robustness, as XC forces are expected to be significant for highly-correlated condensates. The breaking length of NFW is an emergent scale in the otherwise scale-free system of CDM collapse. A similar phenomenon appears to occur with the Gaussian halos. Further, their breaking scale is also invariant across the full range of sampled correlations. Such robustness is unexpected. Further, as self-gravity XC introduces no new scales, Bose halos provide an independent measure of emergent scales among self-gravitating systems. Top-hat halo densities display some XC-induced ordering, sub-dominant to the response to halo spin, especially at the highest tested spin. Top-hat density profiles are characterized by an outer power law of grade $\beta \sim -4.0$ breaking into more mild behavior centered around $\alpha \approx -1.0$ among spinning halos and $\alpha \approx -1.4$ for spin-less. The breaking scales are primarily dependent of spin.

\begin{figure*}
\begin{center}
\includegraphics[width=\textwidth]{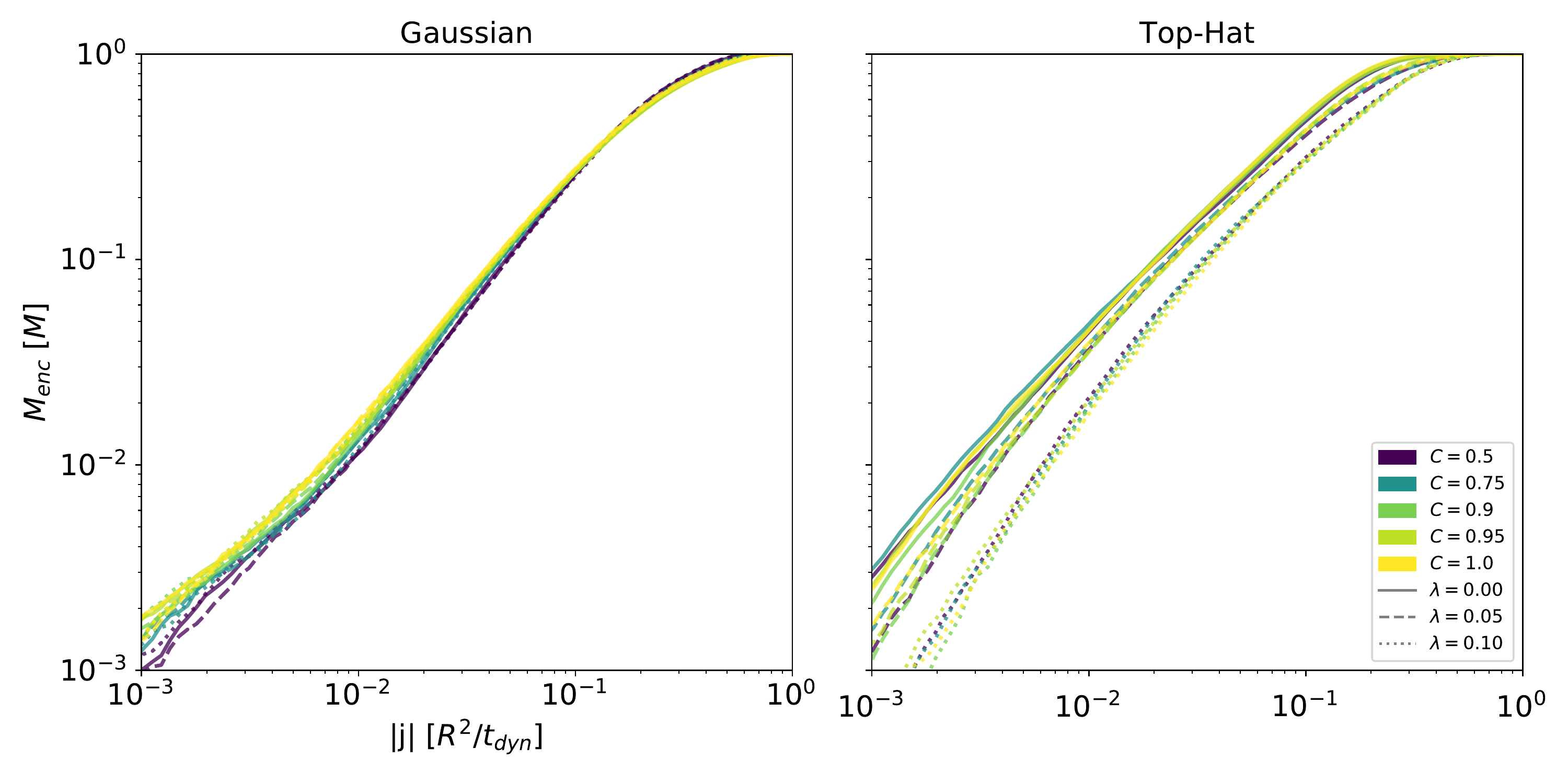}
\caption{Enclosed mass angular momentum profiles of the isolated collapse simulations of Sec.~\ref{SCS}. Gaussian profiles on L, Top-hat profiles on R. Halos were measured after evolving for $10 t_{dyn}$. Profile coloration indicates degree of correlation ranging from classical $C=1.0$ to highly correlated $C=0.5$. Line style indicates level of Peeble's spin $\lambda$ of the halo. }
\label{fig:mencj}
\end{center}
\end{figure*}

Another standard measure of classical infall is provided by the enclosed mass as a function of angular momentum $M_{<} (j)$ \citep{Bullock}, Fig.~\ref{fig:mencj}, and shows marginal but coherent differences across correlation. Gaussian profiles show a relative depletion of low angular momentum material over the first decades, with little sensitivity to spin. The Gauss simulations classically maintain a universal concave up shape, characterized by a breaking roughly at $j_{break} \approx 10^{-2} R^2/t_{dyn}$ between two power laws, as opposed to the single power of cosmological halos found by \citet{Bullock}. The profiles are well ordered and converged in correlation over the high angular momentum region, but begin to diverge at the lower end of the domain. More highly correlated halos are also seen to reach total mass slightly faster than the classical halos. The Top-hat simulations show different profiles, primarily separated by spin, but with additional stratification along XC lines. The shape of Top-hat profiles varies with spin. Spin-less halos are dominated by a single power law of $\alpha \approx -1.25$. Rotating halos are relatively depleted of low angular momentum material and demonstrate concavity throughout the domain.

\begin{figure*}
\begin{center}
\includegraphics[width=\textwidth]{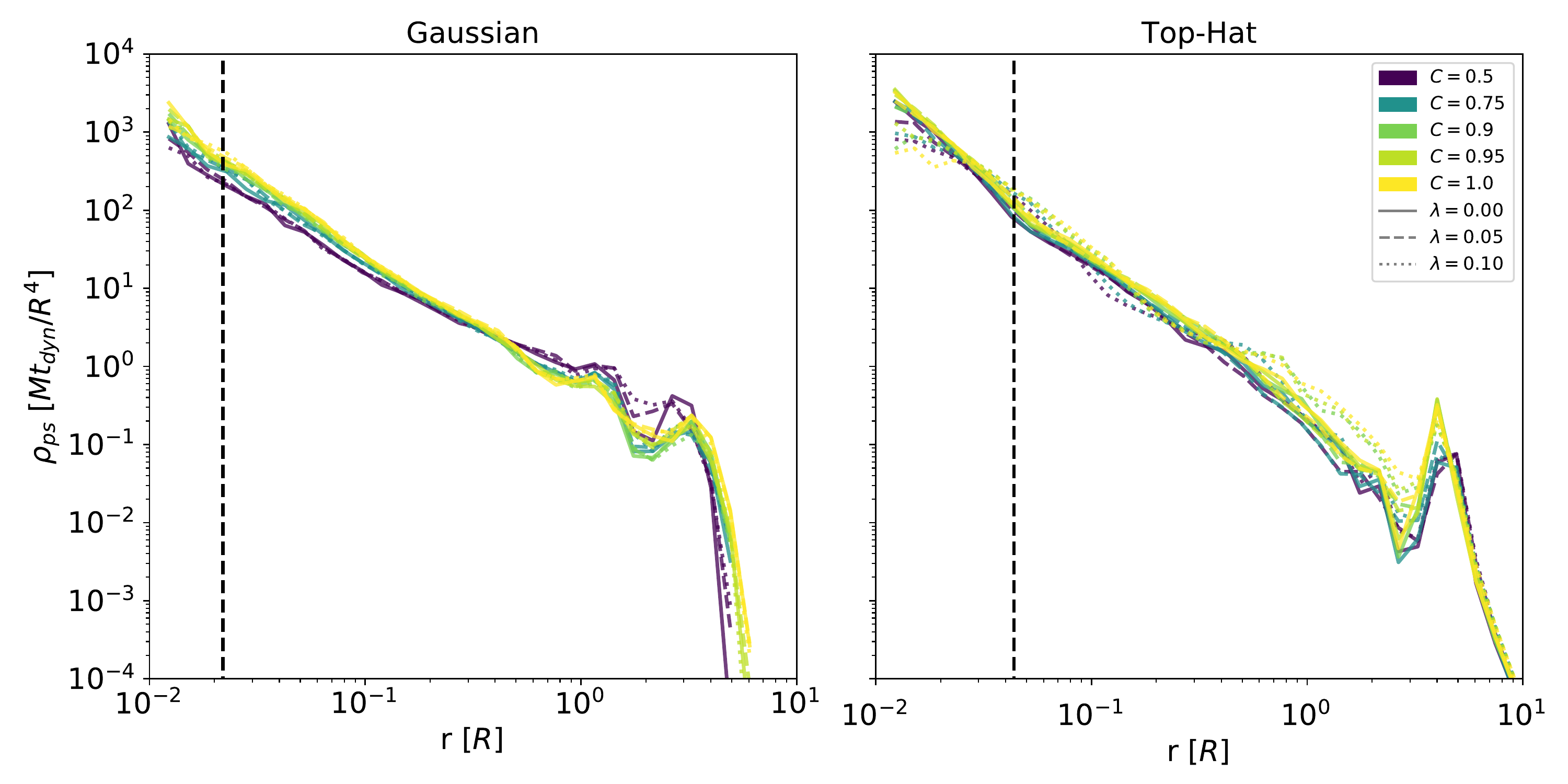}
\caption{Phase space mass density radial profiles of the isolated collapse simulations of Sec.~\ref{SCS}. Gaussian profiles on L, Top-hat profiles on R. Halos were measured after evolving for $10 t_{dyn}$. Profile coloration indicates degree of correlation ranging from classical $C=1.0$ to highly correlated $C=0.5$. Line style indicates level of Peeble's spin $\lambda$ of the halo. Volume in velocity space is measured in the local spherical velocity dispersion. The softening profile's maximum force radius is represented by the black dashed line in each simulation set, below which our confidence in the results are diminished.}
\label{fig:phspcr}
\end{center}
\end{figure*}

\begin{figure*}
\begin{center}
\includegraphics[width=\textwidth]{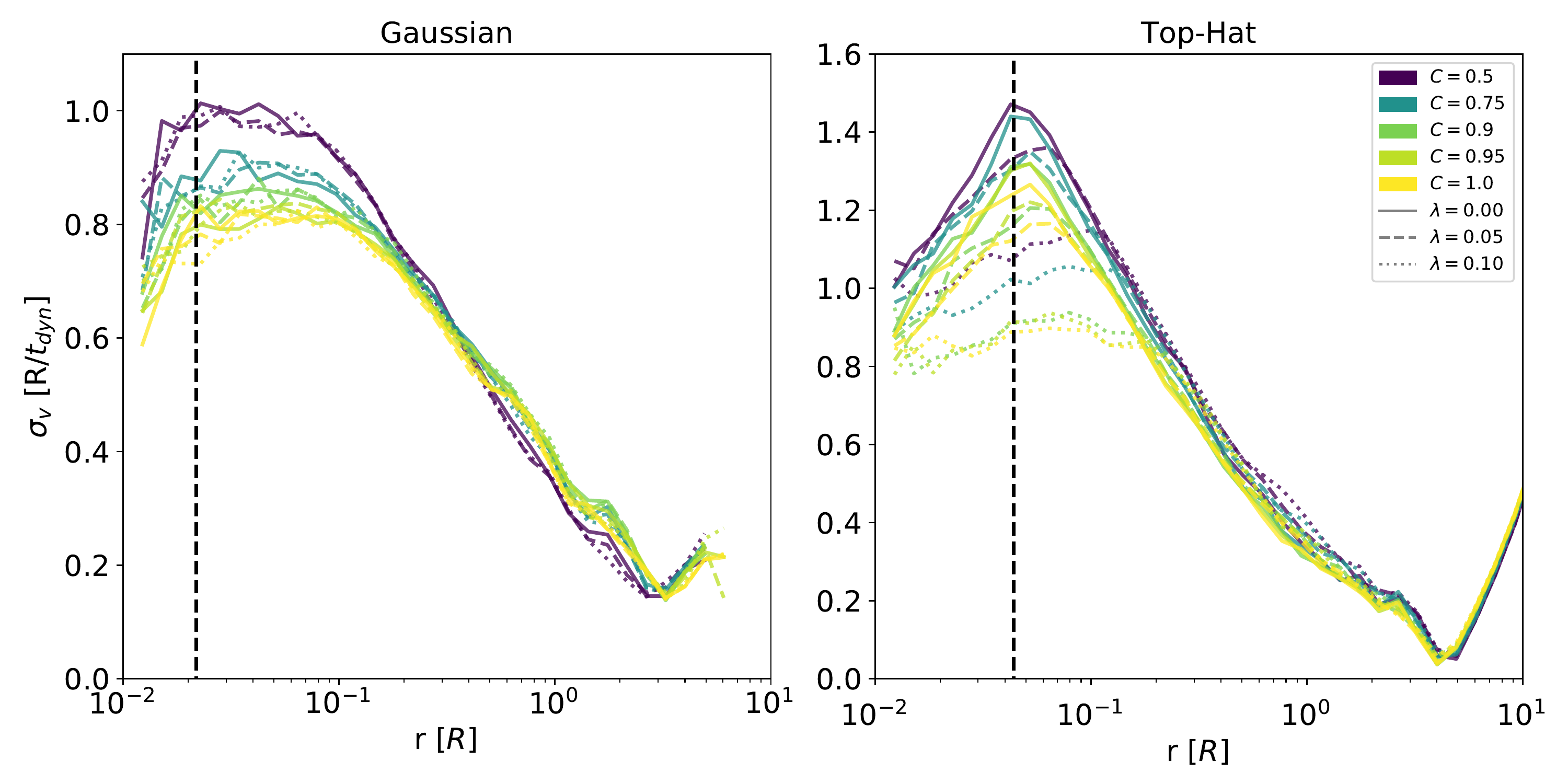}
\caption{Spherical velocity dispersion radial profiles of the isolated collapse simulations of Sec.~\ref{SCS}. Gaussian profiles on L, Top-hat profiles on R. Halos were measured after evolving for $10 t_{dyn}$. Profile coloration indicates degree of correlation ranging from classical $C=1.0$ to highly correlated $C=0.5$. Line style indicates level of Peeble's spin $\lambda$ of the halo. The softening profile's maximum force radius is represented by the black dashed line in each simulation set, below which our confidence in the results are diminished.}
\label{fig:vr}
\end{center}
\end{figure*}

A halo's high-resolution phase-space density profile is expected to show a superposition of falling and rising cold sheets of fluid and their turn-around caustics. Though only the first and perhaps second caustics can be resolved at the outermost radii of these modestly-resolved halos, more concerted changes can be seen in the distribution mean, Fig.~\ref{fig:phspcr}. Gaussian phase space densities show well-consolidated shifts along correlation lines, with virtually all of the spread at inner radii being attributed to the large differences in velocity dispersion, Fig.~\ref{fig:vr}. Note that the profile intersections of Bose halos with their classical analogues, in both the phase-space density and velocity dispersion, occur at the density profile breaking radius. Top-hat halos show differences dominated by correlation for large radius, and spin for small radii. Ordering in correlation exist within each spin class, though there are transitions in the orientation of that ordering, specifically among $\lambda=0.10$ halos. 

\begin{figure*}
\begin{center}
\includegraphics[width=\textwidth]{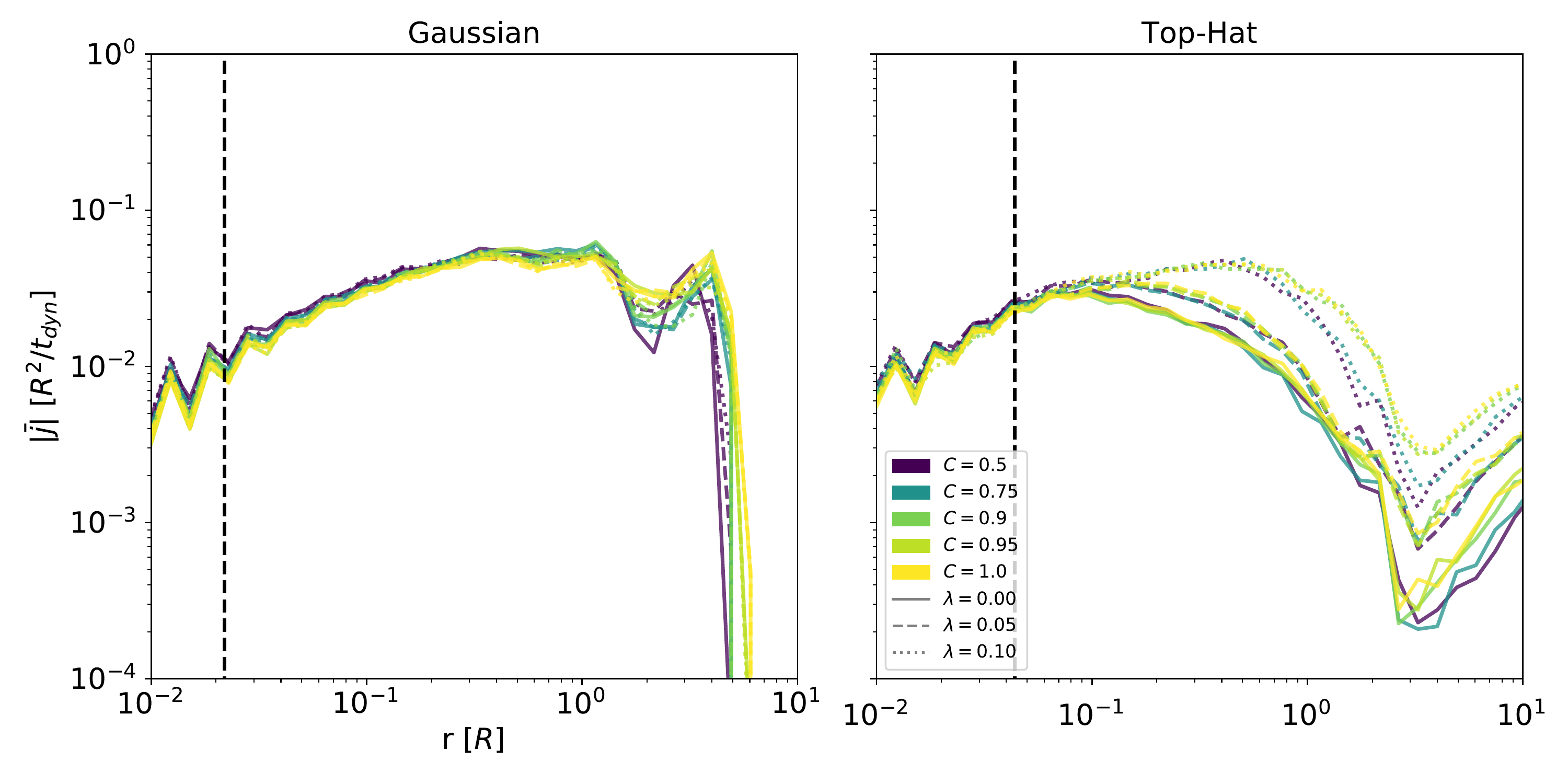}
\caption{Mean magnitude angular momentum radial profiles of the isolated collapse simulations of Sec.~\ref{SCS}. Gaussian profiles on L, Top-hat profiles on R. Halos were measured after evolving for $10 t_{dyn}$. Profile coloration indicates degree of correlation ranging from classical $C=1.0$ to highly correlated $C=0.5$. Line style indicates level of Peeble's spin $\lambda$ of the halo. The softening profile's maximum force radius is represented by the black dashed line in each simulation set, below which our confidence in the results are diminished.}
\label{fig:jr}
\end{center}
\end{figure*}

Radial profiles of angular momentum and velocity dispersion display new structural features more readily than density, Figs.~\ref{fig:vr},\ref{fig:jr}. Gaussian halos, as mentioned above, show clear augmentation of velocity dispersion with correlation, with transitions  from amplification to compression occuring at the breaking radius, and are fairly insensitive to spin. Top-hat halos also show clear amplification of dispersion, but no transition. Shapes of the Top-hat velocity dispersion profiles change notably with respect to spin. Spin-less halos show a symmetric low radius dispersion peak in logarithmic scaling near the softening length. Spun halos flatten this peak and shift its diminished maximum to higher radii, where it converges with the more weakly spun halos.  Sample angular momentum differences are less pronounced, possibly due to the relatively conserved status of angular momentum in a system with spontaneously broken rotational symmetry. Most particle angular momentum evolution is induced by the ROI symmetry breaking feature. Correlations consistently produce a first higher and then lower mean angular momentum value as one moves out in radius for Gaussian halos, again transitioning at the breaking radius. Top-hat collapses behave similarly at small radii, eventually branching according to spin. Each spin branch exhibits its own correlation substructure.

\begin{figure*}
\begin{center}
\includegraphics[width=\textwidth]{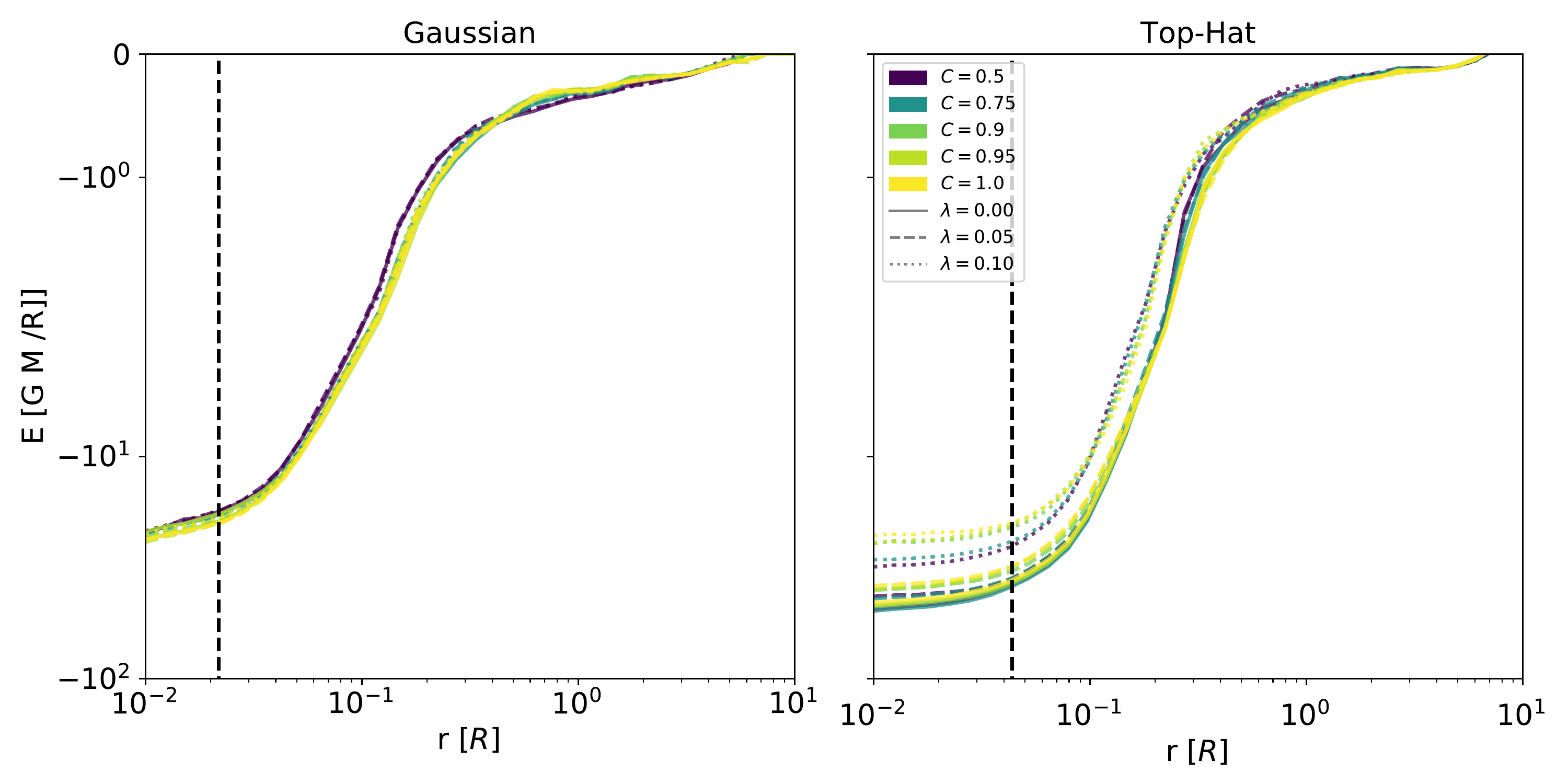}
\caption{Classical binding energy radial profiles of the isolated collapse simulations of Sec.~\ref{SCS}. Gaussian profiles on L, Top-hat profiles on R. Halos were measured after evolving for $10 t_{dyn}$. Profile coloration indicates degree of correlation ranging from classical $C=1.0$ to highly correlated $C=0.5$. Line style indicates level of Peeble's spin $\lambda$ of the halo. The softening profile's maximum force radius is represented by the black dashed line in each simulation set, below which our confidence in the results are diminished.}
\label{fig:Er}
\end{center}
\end{figure*}

Classical binding energy per-particle is another near integral of motion of the classical virialized halo, Fig.~\ref{fig:Er}. Classical energy density is only expected to be an integral of the motion once quasi-equilibrium is satisfied, and then only for classical samples with an effective Hamiltonian of canonical type. Bose halo fluid elements are not expected to have integral energy, though the system's total energy can be shown to be integral. Matching of energy profiles is well-consolidated over correlation for multiple decades in radius among Gaussian halos. Bose halos show up to a factor of two difference in per-particle energy over classical halos between $r=0.1-0.2R$. Correlated halos show deeper binding energy at these large radii. Bose Gaussian halos are also smoother in profile. The more consistent and slower scaling of energy among Bose halos at large radii leads to a transition to shallower energies within the breaking radius. Top-hat halo energy profiles are also seen to be well ordered, and, within their spin sets, can show a much higher separation in energy than Gaussian profiles. This is expected after observing the effect of spin to lower the density profile, resulting in a shallower potential well for classical halos.

The shape of the halo beyond spherical or cylindrical symmetry is important to understanding the tangential dynamical mechanisms at play. Halo triaxalities are found to be quite dynamic throughout the virial volume of a halo. Among Gaussian halos, triaxalities generally range from moderately oblate to spherical at inner radii, increasing in prolate-ness until slightly outside the softening length $r \sim 0.05R$ and then relaxing to a modest prolate state for the remainder of the virialized volume. Bose halos show a slightly higher tendency for prolate shape than classical halos at outer radii. While there are other differences in shape observed among the classical and Bose halos, their behavior with respect to level of correlation and spin is not perfectly clear. Top-hat triaxalities show a steady trend among spin-less halos from a spherical shape at inner radii to a more prolate state of $T = 0.7-0.8$ farther out. Spun halos are far more varied, starting with oblate or spherical shape at inner radii just as the Gaussian halos, rising to near total prolate-ness before falling to a spherical or oblate state at large radii. Again the role of correlation for these halos remains unclear.

\begin{figure*}
\begin{center}
\includegraphics[width=\textwidth]{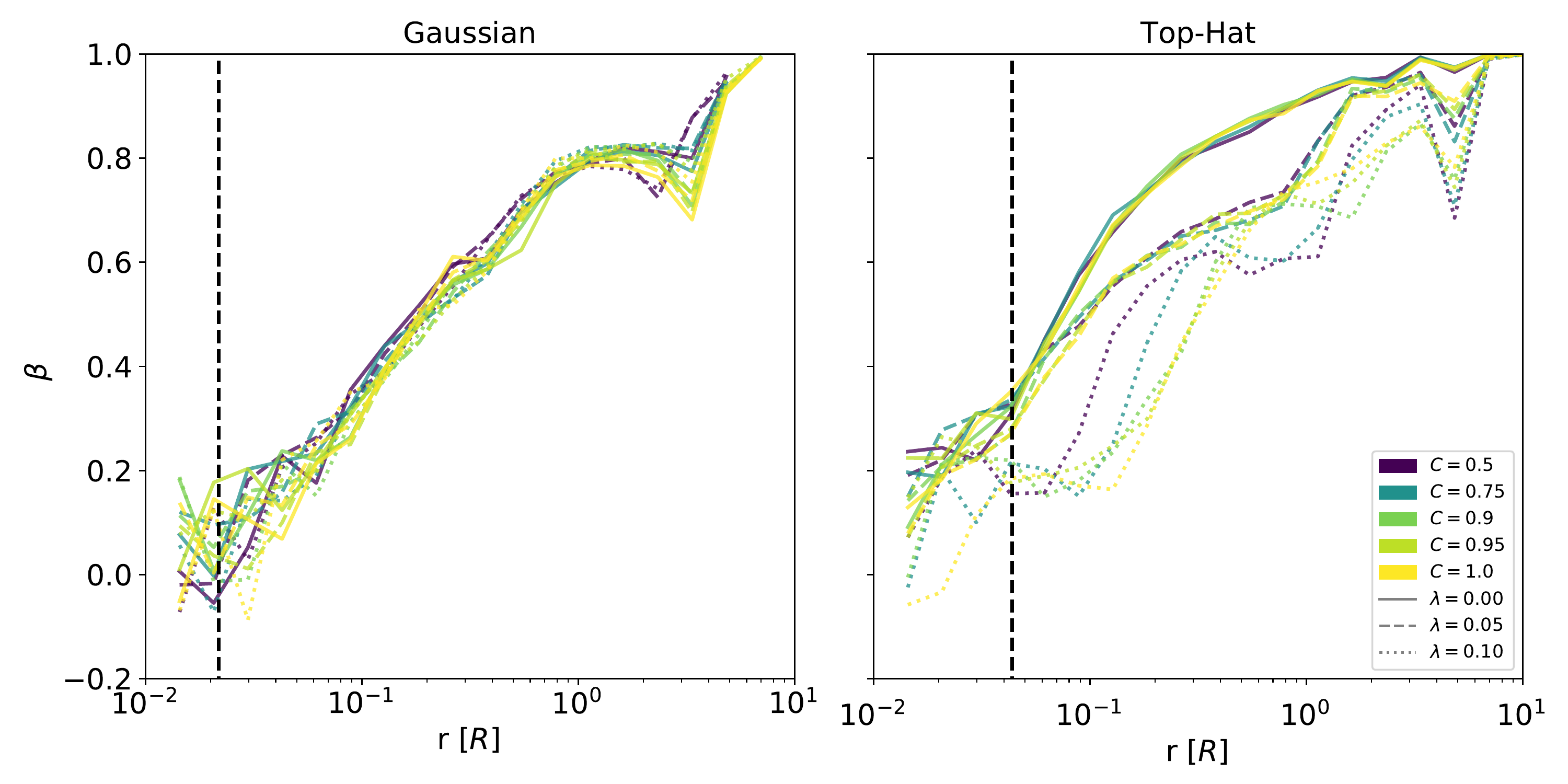}
\caption{Velocity anisotropy radial profiles of the isolated collapse simulations of Sec.~\ref{SCS}. Gaussian profiles on L, Top-hat profiles on R. Halos were measured after evolving for $10 t_{dyn}$. Profile coloration indicates degree of correlation ranging from classical $C=1.0$ to highly correlated $C=0.5$. Line style indicates level of Peeble's spin $\lambda$ of the halo. The softening profile's maximum force radius is represented by the black dashed line in each simulation set, below which our confidence in the results are diminished.}
\label{fig:betar}
\end{center}
\end{figure*}

The local shape of a halo's speed distribution, or anisotropy, is nearly as robust to XC interactions as the mass density profile, Fig.~\ref{fig:betar}. Weakly spun halos generally show a velocity anisotropy in favor of motion in the radial direction as a consequence of their initial near-radial infall, though the collection of orbits tend towards ergodicity ($\beta=0$) at inner radii. Much of the virial volume among Gaussian halos shows little to no discernible Bose structure. Only at the edge of the virial volume do correlation features stratify, with more correlated halos softening the hard break in anisotropy as one crosses the virial boundary. A small shift in the virial break is also observed for the most highly correlated spinning halos. Top-hat halos do not all reach ergodicity. Each spin state has its own preferred profile shape. Correlation dependence is difficult to distinguish among the lower spin states, though some coherent shifts are seen between the most highly spun halos.

\begin{figure*}
\begin{center}
\includegraphics[width=\textwidth]{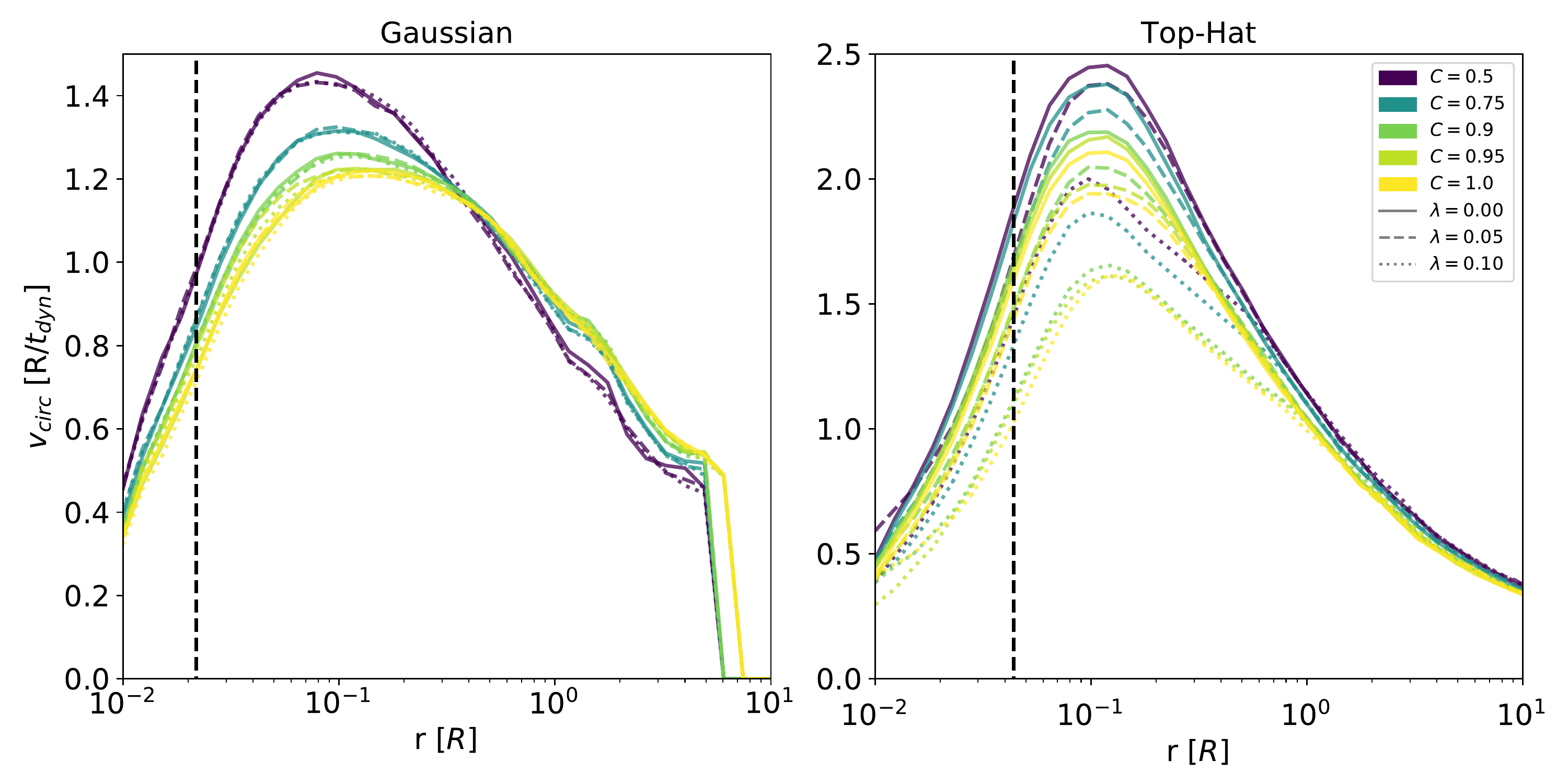}
\caption{Circular orbit speed radial profiles of the isolated collapse simulations of Sec.~\ref{SCS}. Gaussian profiles on L, Top-hat profiles on R. Halos were measured after evolving for $10 t_{dyn}$. Profile coloration indicates degree of correlation ranging from classical $C=1.0$ to highly correlated $C=0.5$. Line style indicates level of Peeble's spin $\lambda$ of the halo. The softening profile's maximum force radius is represented by the black dashed line in each simulation set, below which our confidence in the results are diminished.}
\label{fig:circvr}
\end{center}
\end{figure*}

Circular orbit speed profiles serve as a proxy for experienced force-per-particle, and quantify much of the virialized halo dynamics, Fig.~\ref{fig:circvr}. Each profile follows an expected arc as the density steepens with rapidly increasing enclosed mass before leveling out and dropping to the edges after new mass has been depleted. The differences in force over the correlated curves are large. The most highly correlated $C=0.5$ experience up to a $70\%$ higher than classical at peak force, more than twice classical at the center, and lower than classical force on the outer reaches in the Gaussian halos. The force augmentation translates to differences in the circular speed curves of up to $45\%$ at any given radius, and a $20\%$ increase in maximum speed. Bose Top-hat halos also see a boost in force of up to $70\%$ and an increase of rotational speed by over $30\%$. Again, it is surprising that higher acceleration of sample points does not much change the spatial structure of the system. The crossing between stronger than classical and weaker than classical force occurs near the breaking radius among Gaussian halos, coinciding with similar transitions in the velocity dispersion and classical energy. Bose Top-hat halos experience higher than classical force throughout.

\begin{figure}
\begin{center}
\includegraphics[width=9cm]{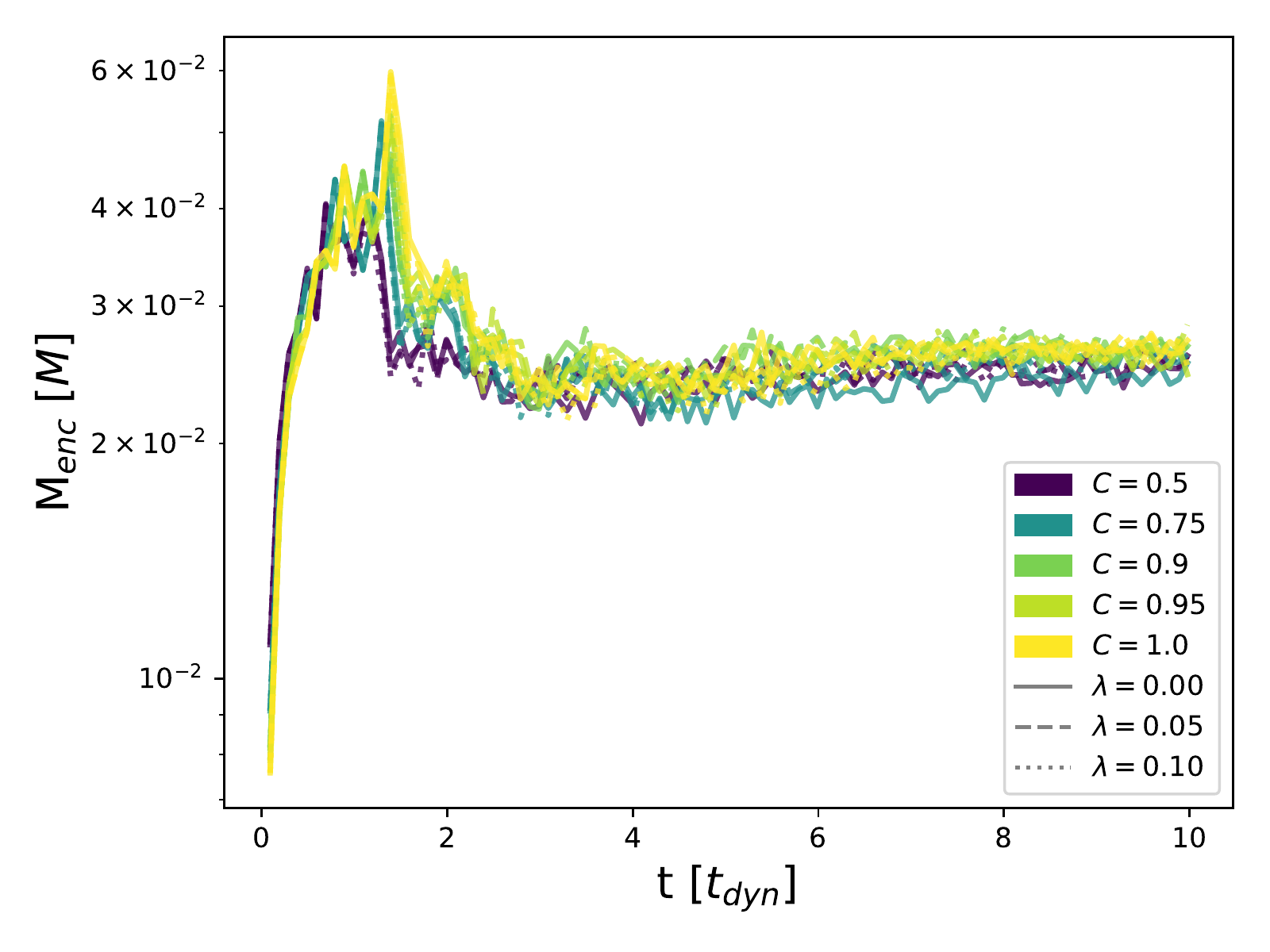}
\caption{Mass within central $r=0.05R$ of halo over time of Gaussian isolated collapse simulations of Sec.~\ref{SCS}. Profile coloration indicates degree of correlation ranging from classical $C=1.0$ to highly correlated $C=0.5$. Line style indicates level of Peeble's spin $\lambda$ of the halo. }
\label{fig:mtracking}
\end{center}
\end{figure}

Several more Bose features can be seen by following the central distribution of matter through infall, Fig.~\ref{fig:mtracking}. An increase in collapse rapidly to first infall is seen, up to $\sim 20\%$ among Gaussian initial conditions at $C=0.5$ and $\sim 30\%$ faster infall for Top-hat. Faster collapse is first demonstrated in LQR and is expanded on in Appx.~\ref{SShell}. Chaotic dynamics and violent relaxation of the halo's most extreme phase characterize the evolution immediately following first shell crossing. Violent relaxation damps after many local crossing times, or several system dynamical times, into a more settled quiescent phase of evolution by $\sim 4 t_{dyn}$. The chaotic phase of Bose simulations is seen to pass more quickly, with the most correlated systems entering quiescent phase more than one dynamical time earlier. Halo spin seems to have only a marginal impact on the features of the dynamical central mass.

\subsection{Global Distributions and Law-Breaking Features}

\begin{figure*}
\begin{center}
\includegraphics[width=\textwidth]{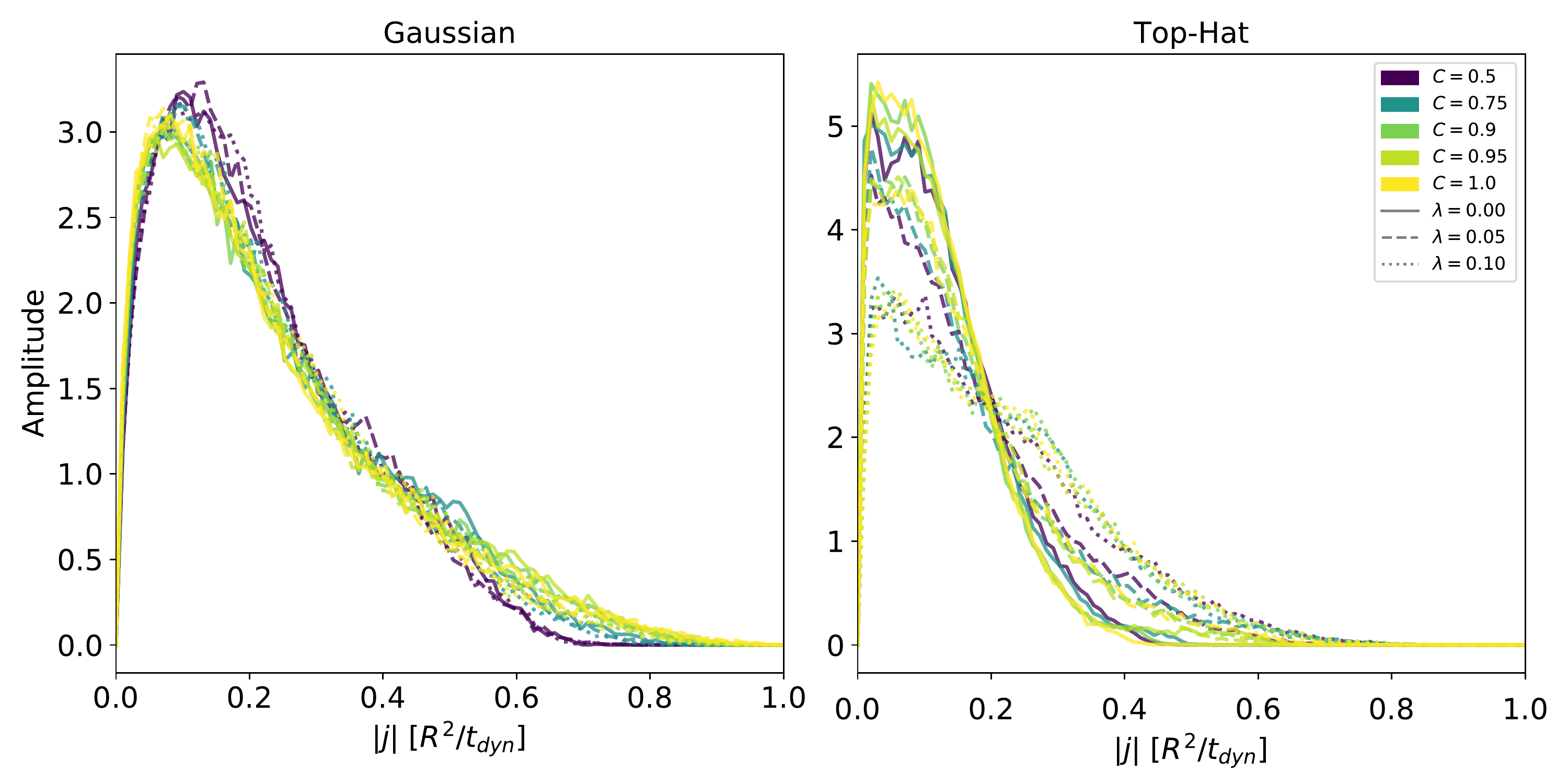}
\caption{Magnitude angular momentum distribution function of the isolated collapse simulations of Sec.~\ref{SCS}. Gaussian profiles on L, Top-hat profiles on R. Halos were measured after evolving for $10 t_{dyn}$. Profile coloration indicates degree of correlation ranging from classical $C=1.0$ to highly correlated $C=0.5$. Line style indicates level of Peeble's spin $\lambda$ of the halo. }
\label{fig:jdistr}
\end{center}
\end{figure*}

More direct projections of the final distribution function reveal other novel Bose structures. Magnitude angular momenta among Gauss simulations show organization along correlation for their entire domain, ending with all spins of a single correlation having the same support, Fig.~\ref{fig:jdistr}. The shortening of angular momentum support with correlation may be an indication of the XC interaction as a means of maintaining compactness. Top-hat halos expectedly divide firstly along spin, but also show minor correlation organization, most notably at $\lambda=0.10$. 

\begin{figure*}
\begin{center}
\includegraphics[width=\textwidth]{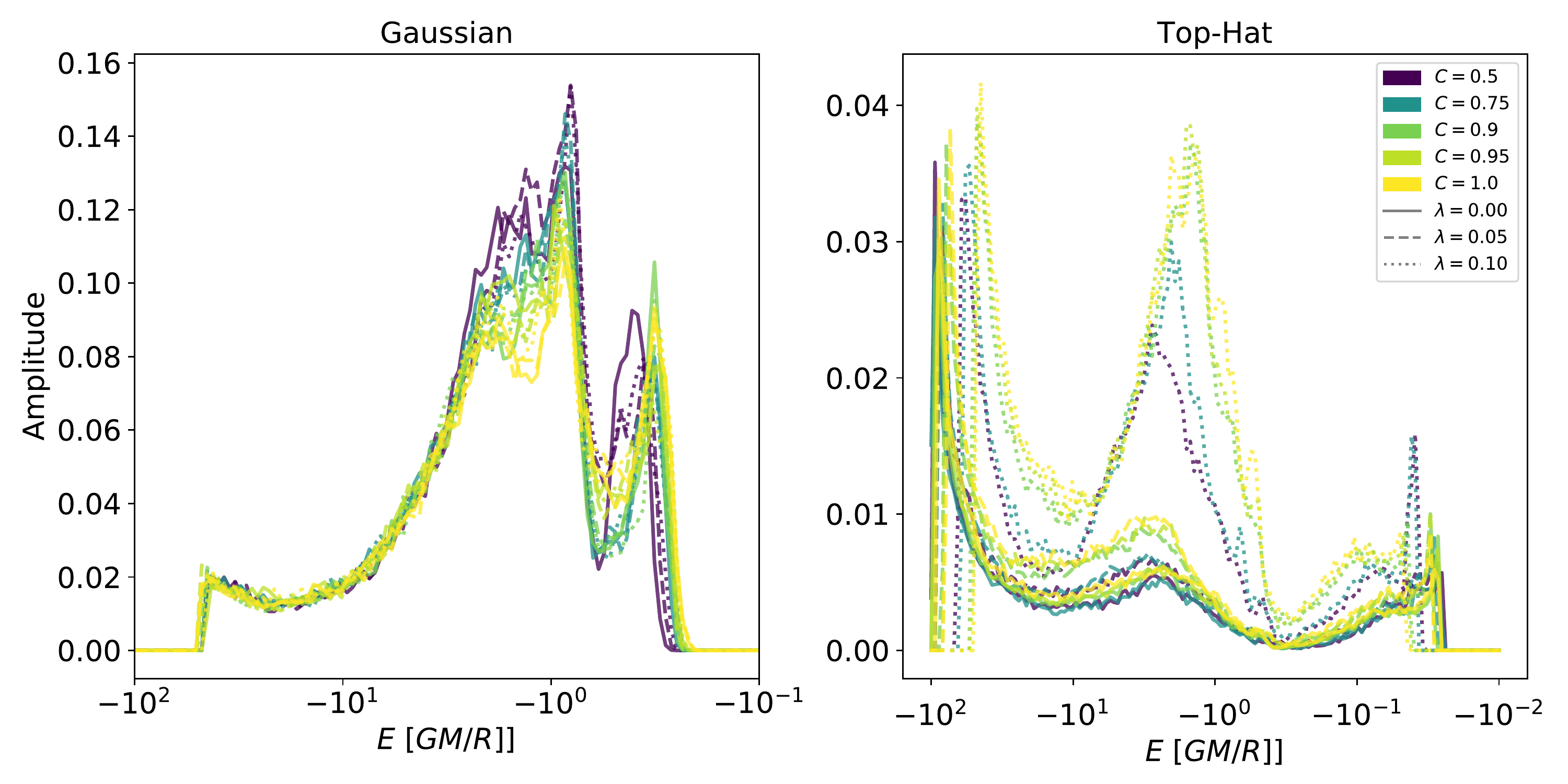}
\caption{Classical binding energy distribution function of the isolated collapse simulations of Sec.~\ref{SCS}. Gaussian profiles on L, Top-hat profiles on R. Halos were measured after evolving for $10 t_{dyn}$. Profile coloration indicates degree of correlation ranging from classical $C=1.0$ to highly correlated $C=0.5$. Line style indicates level of Peeble's spin $\lambda$ of the halo. Domain is restricted to classically bound states.}
\label{fig:Edistr}
\end{center}
\end{figure*}

Classical binding energy distributions are also well organized, showing concerted changes in substructure, Fig.~\ref{fig:Edistr}. Gauss halos, all with approximately the same minimum potential, show uniform growth for the most tightly bound objects. Populations at higher energy begin to diverge in shape according to correlation. Classical halos are seen to have several highly populated regions above $E=-3 GM/R$, with a central peak at $E \approx -1 GM/R$. The highest energy of these peaks and the adjacent low energy trough become depleted with XC forces through $C=0.75$, after which the peak begins to shift to lower energies and re-establishes itself to some degree among $C=0.5$ halos. Correlated distributions also shift their lower energy spectra into a central consolidated feature, still centered at $E=-1 GM/R$. Top-hat energy structure is dominated by spin, but also displays obvious signs of correlation. Much of the observable correlated structure is in the form of more prominent high-energy bands, less prominent middle-energy structure, and lower potential floors. Classically unbounded objects are not displayed in these distributions.

Classical halos, evolved well into quasi-equilibrium, obey a simple classical energy constraint: bodies with kinetic energy above the asymptotic potential binding are unbounded to the system and will separate themselves from the system on the order of the crossing time. As our halos are several crossing times into their virialized states, there should be no unbounded bodies remaining in the classical halos. Bose fluids do not necessarily obey this condition as they are subject to an additional interaction not captured by the gravitational potential. Regular violation of the classical energy condition is another potential marker of unique Bose structure.

\begin{figure*}
\begin{center}
\includegraphics[width=\textwidth]{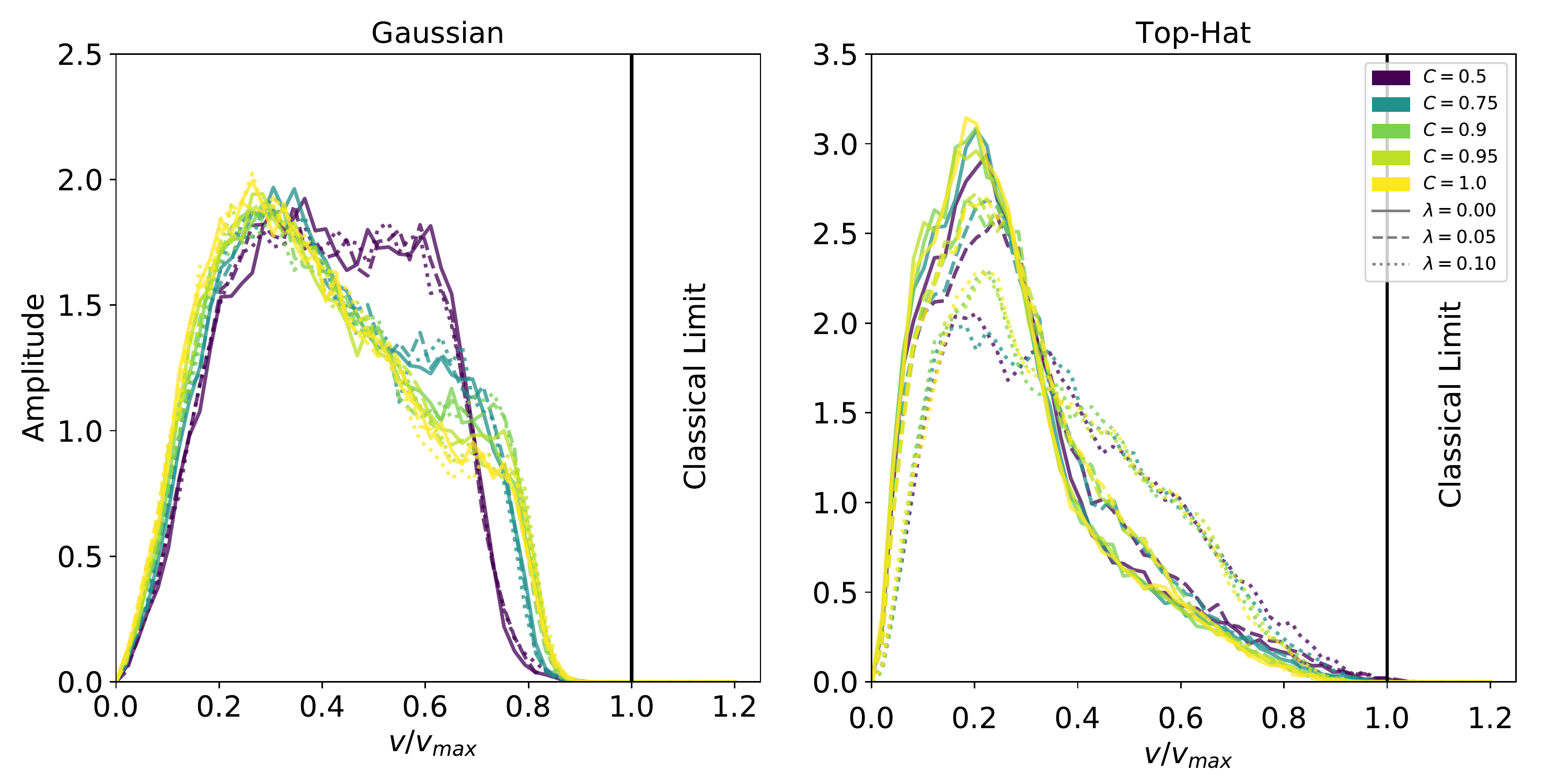}
\caption{Distribution functions of speed fraction over classical gravitational potential ($v/v_{max} = v/\sqrt{2 |\bar{\Phi}|}$) of Top-hat isolated collapse simulations of Sec.~\ref{SCS}. Gaussian profiles on L, Top-hat profiles on R. Halos were measured after evolving for $10 t_{dyn}$. Distributions are taken from within $r=0.4R$ for Gaussian simulations and $r=1.0R$ for Top-hat. Profile coloration indicates degree of correlation ranging from classical $C=1.0$ to highly correlated $C=0.5$. Line style indicates level of Peeble's spin $\lambda$ of the halo. Sample points lying outside of the classical limit occur only with halos of correlation $C=0.75$ or better, and amount to about one tenth of a percent of the sampled mass.}
\label{fig:vfracdistr}
\end{center}
\end{figure*}

 Bose halos show a concerted shift of their velocity distribution with respect to the classical binding limit $v_{\max} = \sqrt{2 \bar{\Phi}}$, Fig.~\ref{fig:vfracdistr}. Angular momenta, speeds, or other kinetic quantities have their own form of the classical energy constraint. Speed is shown here, since it produces a larger effect than say angular momentum, which is subject to equi-partition of velocity in-so-far as a virialized structure can manage it. Sampling of halos is taken where the fractional and overall XC forces are highest, within the breaking radius $r_{G} = r_{break} = 0.4 R$ for Gauss halos and Top-hat halos inside of $r_{TH} = 1.0R$. Classical structures observe the binding energy limit in both sets of collapse. The default shape among Gaussian halos is a single low speed peak at $v/v_{max} \approx 0.25$, followed by a smooth tail that turns steeply at $v/v_{max} \approx 0.8$. Gaussian Bose halos have slower tails at both the high and low speed ends, with the high-speed tail extending slightly beyond the classical cutoff, and with the bulk of the distribution shifting slowly to the double-lobed plateau seen in $C=0.5$ halos. Distribution centers of mass move little with correlation, staying close to $\braket{v/v_{max}} \approx 0.4$. Top-hat halos are closer to a thermal distribution in shape, with long high-speed tails. Each spin group has its own class of shapes. Distributions within each spin change with correlation, becoming depleted at low speed in favor of elongating the high-speed tail beyond the the classical limit. Double peaks are also observed for Top-hat halos of high spin and correlation.

The binding-condition-violating sub-population among Top-hat simulations is a small but predictable group. Spin of the halo does not play a significant role in the size of the sub-population. Boosted samples are present throughout the halo volume, save for the edge of the outer most orbiting sheets. The lack of energy-condition-violating bodies in Gaussian halos is perhaps due to the generally gentler infall of the normal profile, resulting in less work done on individual particles. There are no unbound bodies at any correlation among Guassian halos.

Measuring such changes in the DM speed distribution of an isolated halo is difficult to accomplish with baryons. Stellar populations are not directly subject to the modified gravity from XC, only the total Newtonian gravitational potential. Axion searches sensitive to a halo's local energy distribution may be capable of resolving differences in the kinetic makeup of a Bose halo, given sufficient integration time. Another means to observe the explicit DM motion may come from the motions and organization of satellite galaxies and other halo substructure.

The resolution limit of a single snapshot prevents much more from being said on the nature of halo structure. The next subsection introduces a stacking method in order to expand both the dimension and resolution of measurements.

\subsection{Fine Structure and Orbital Actions}
\label{finestruct}

Each halo is well settled into quasi-equilibrium after ten dynamical times. A sub-sampling of snapshots after virialization and prior to relaxation can be organized into a stack of phase-mixed quasi-independent snapshots of the halo. The elements of such a stack can be co-added, effectively increasing the number of samples in the Monte-Carlo. Extending several of the $S(,,)$ simulations by another multiple dynamical times past virialization can also provide samples of particle orbits within the virial radius, allowing for an orbital analysis of the halos. Stacking and orbit integration together create the most resolved pictures of the current analysis. The simulations chosen for extension are of Gaussian shape, of all sampled spins, of correlations $\{0.5,0.75,0.9,1.0\}$, and are prolonged by $2 t_{dyn}$. 

True integral actions in a triaxial problem are difficult to construct, despite the limited number needed to span a equilibrated DF \citep{BT2008}. The three desired integrals are approximated here with three accessible actions. Two of the actions are tangential
\begin{align}
j_{\phi} &= \bmath{j} \cdot \hat{\bmath{z}} \\
j_{\theta} &= | \bmath{j} - j_{\phi} \hat{\bmath{z}}| 
\end{align}
and are expected to be approximately integral. The third and final independent action of the set is taken to be the conjugate to radius via the Fourier transform, $j_r$, providing harmonic weights over the two dynamical times of sampling via $\tilde{r}(j_r)$. Note that the $\hat{\bmath{z}}$ vector used in the definition of $j_{\phi}$ aligns with the direction of net spin for halos with $\lambda>0.0$. For spin-less halos, $\hat{\bmath{z}}$ is chosen as the direction of greatest angular momentum dispersion, remaining consistent with the spun halo orientation in the presence of an ROI. Classical energy can also be used as a substitute for the radial action.

\begin{figure*}
\begin{center}
\includegraphics[width=\textwidth]{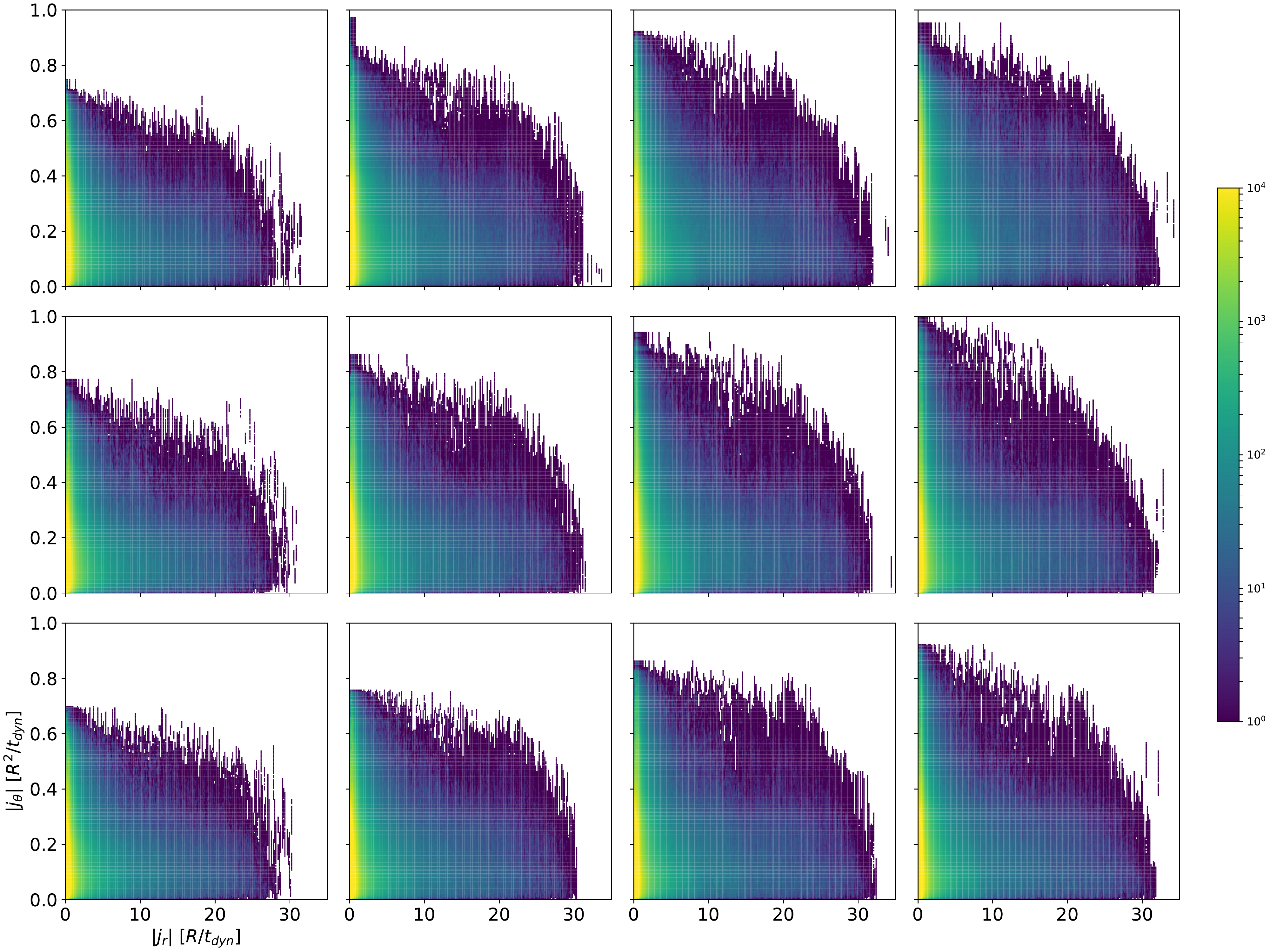}
\caption{Radial and polar orbital action distributions of Gaussian isolated collapse simulations. panels are organized by correlation (column-wise: C=0.5,0.75,0.9,1.0 left to right) and spin (row-wise: $\lambda$=0.0,0.05,0.1 top to bottom). Samples are taken from $100$ equally-spaced frames during $10 - 12 t_{dyn}$. Sample points are taken such that a particle's weights are given by the outer product of $j_{\theta}$ and the radial auto-correlation power spectrum.}
\label{fig:thetaraction}
\end{center}
\end{figure*}

Probability distribution functions in higher dimensions provide a broader view of the distribution landscape, and are used to great effect here in light of the higher count statistics. Action distributions show some concerted changes in global structure, Fig.~ \ref{fig:thetaraction}. Both transverse and radial actions show more condensed support, much like the one-dimensional angular momentum action of Section~\ref{GF}. Some finer structure changes are also visible in the transverse actions, perhaps due to their relation with the ROI. The compression of the radial action with correlation implies the excitement of fewer high frequency radial modes. The consistent reduction in width along all three actions in Bose halos is similar to the condensation of motion that has been seen in other degrees of freedom. The action supports vary some with coherent solid-body spin for classical collapse, but spin susceptibility is somewhat subdued when the XC interaction become sizable.

\begin{figure*}
\begin{center}
\includegraphics[width=\textwidth]{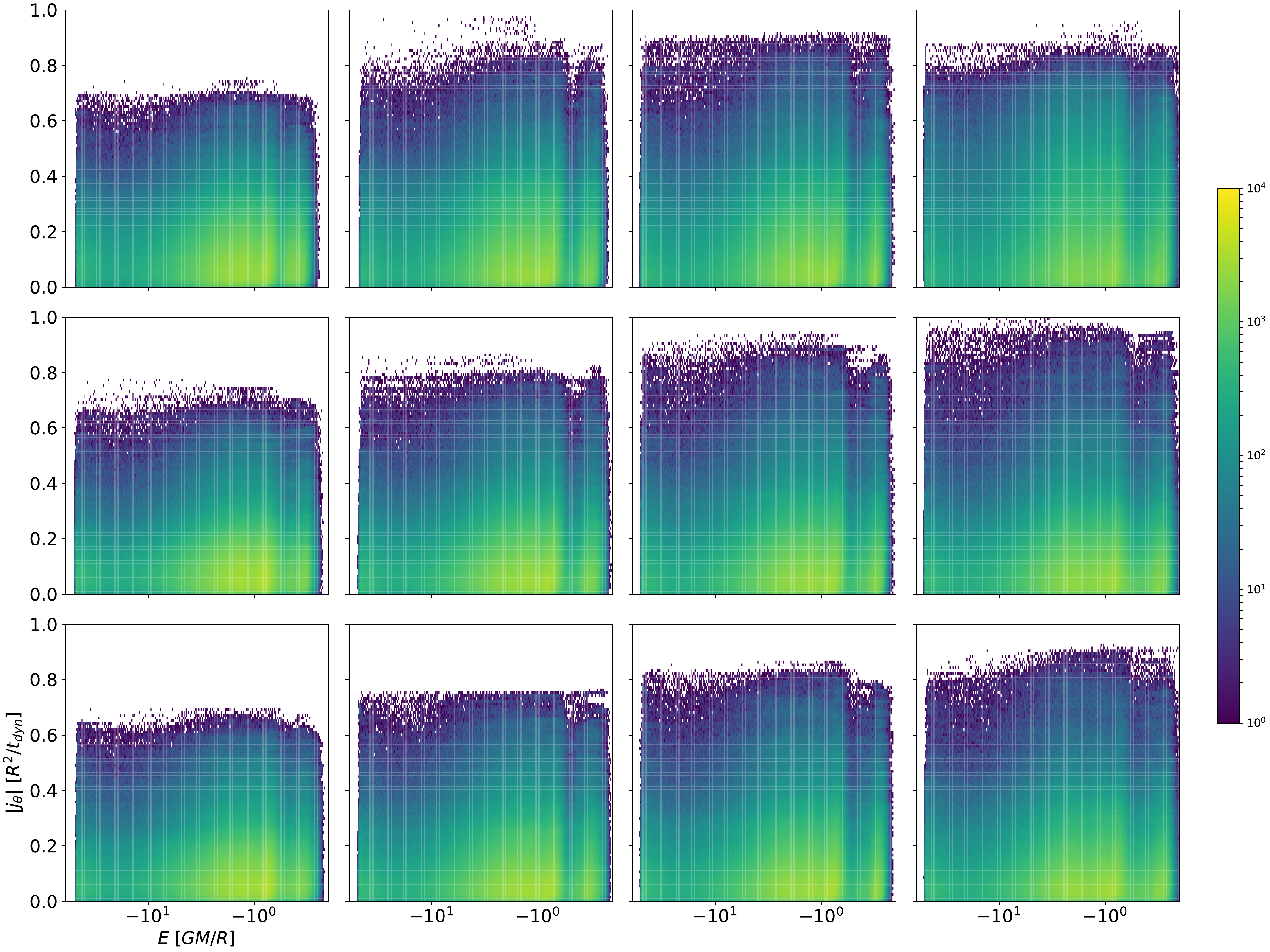}
\caption{Classical binding energy and polar orbital action distributions of Gaussian isolated collapse simulations. panels are organized by correlation (column-wise: C=0.5,0.75,0.9,1.0 left to right) and spin (row-wise: $\lambda$=0.0,0.05,0.1 top to bottom). Samples are taken from $100$ equally-spaced frames during $10 - 12 t_{dyn}$. Sample points are taken such that a particle is given a $j_{\theta}$ and an independent classical energy at each frame. Note that the projection along energy give a higher-resolution version of Fig.~\ref{fig:Edistr}.}
\label{fig:Ethetaaction}
\end{center}
\end{figure*}

The energy spectra over transverse actions show bands of relative over-population or under-population, features present in the single snapshot, and becoming visibly reinforced in the amalgam, Fig.~\ref{fig:Ethetaaction}. Some of these features may be related to back-splash shells from first shell crossings \citep{Mansfield2017} or other processes, especially deeper in the potential well. The finer lines show a great deal of variablilty across correlation, such that it is difficult to see how one may continuously deform into another, though they remain quite consistent across spins. A small but pronounced subset of spectral lines are visible at each correlation, like the main features of the one-dimensional energy distribution of Fig.~\ref{fig:Edistr}. Little deformation of the lines occur with respect to transverse action.

\begin{figure*}
\begin{center}
\includegraphics[width=\textwidth]{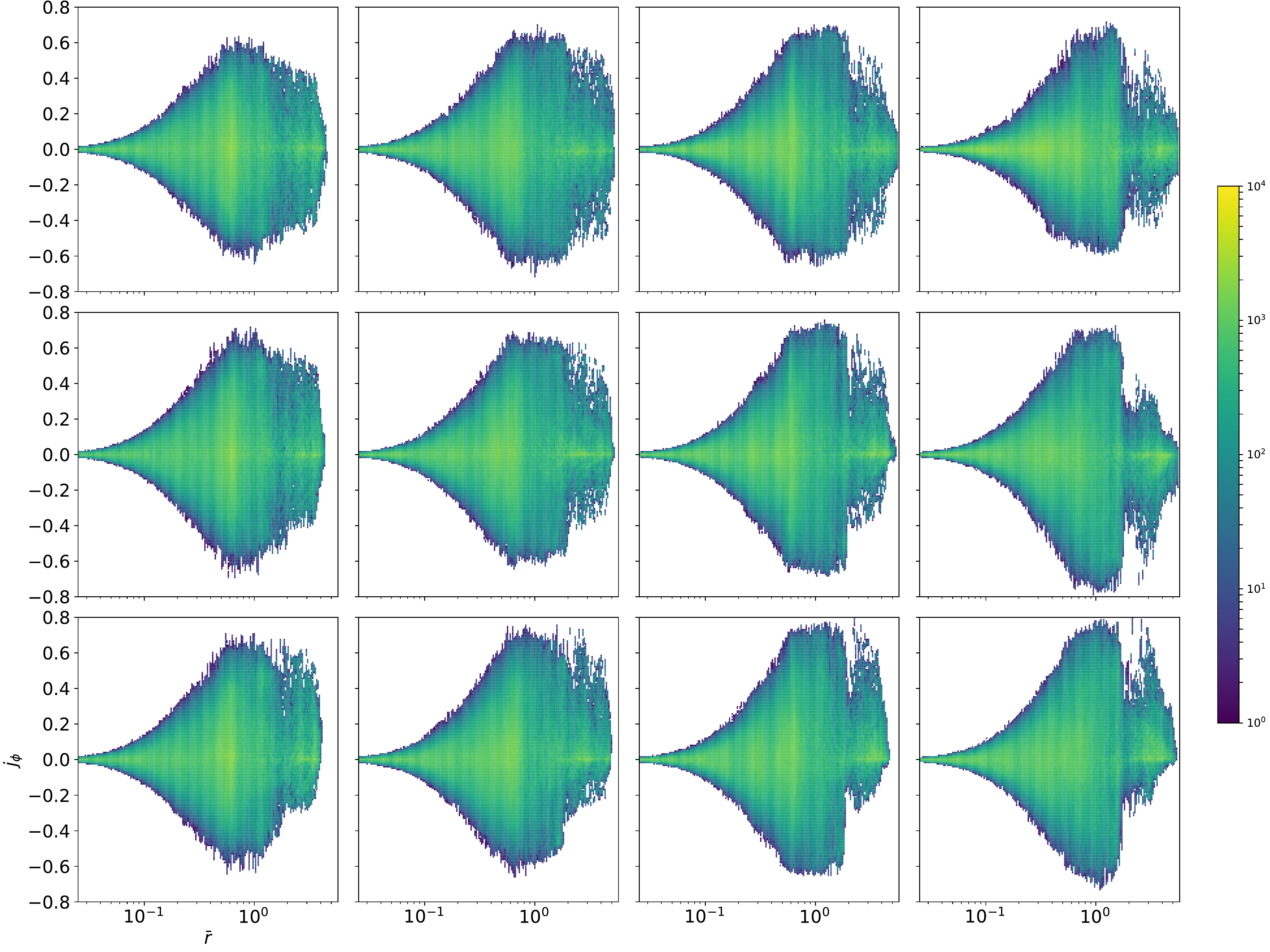}
\caption{Azimuthal orbital action and mean orbit radius distributions of Gaussian isolated collapse simulations. panels are organized by correlation (column-wise: C=0.5,0.75,0.9,1.0 left to right) and spin (row-wise: $\lambda$=0.0,0.05,0.1 top to bottom). Samples are taken from $100$ equally-spaced frames during $10 - 12 t_{dyn}$. Sample points are taken such that each  particle is given one mean radius and an independent $j_{\phi}$ at each frame.}
\label{fig:rmeanphiaction}
\end{center}
\end{figure*}

How the actions change throughout the halo, which shall be represented here by the particle's mean orbital radius, also hold signatures of Bose physics. Tangential actions show a distinct transition between the virialized and un-virialized regions, Figs.~\ref{fig:rmeanphiaction}. Classical halos show a well-filled spectra within the virial radius, then transition discontinuously to one of relatively sparse population. The transition across the virial boundary becomes weaker for more highly correlated halos, producing a more continuous and coherent conversion. Fine structure is not well resolved along this action due to the near continuous spectra available in the transverse directions.

\begin{figure*}
\begin{center}
\includegraphics[width=\textwidth]{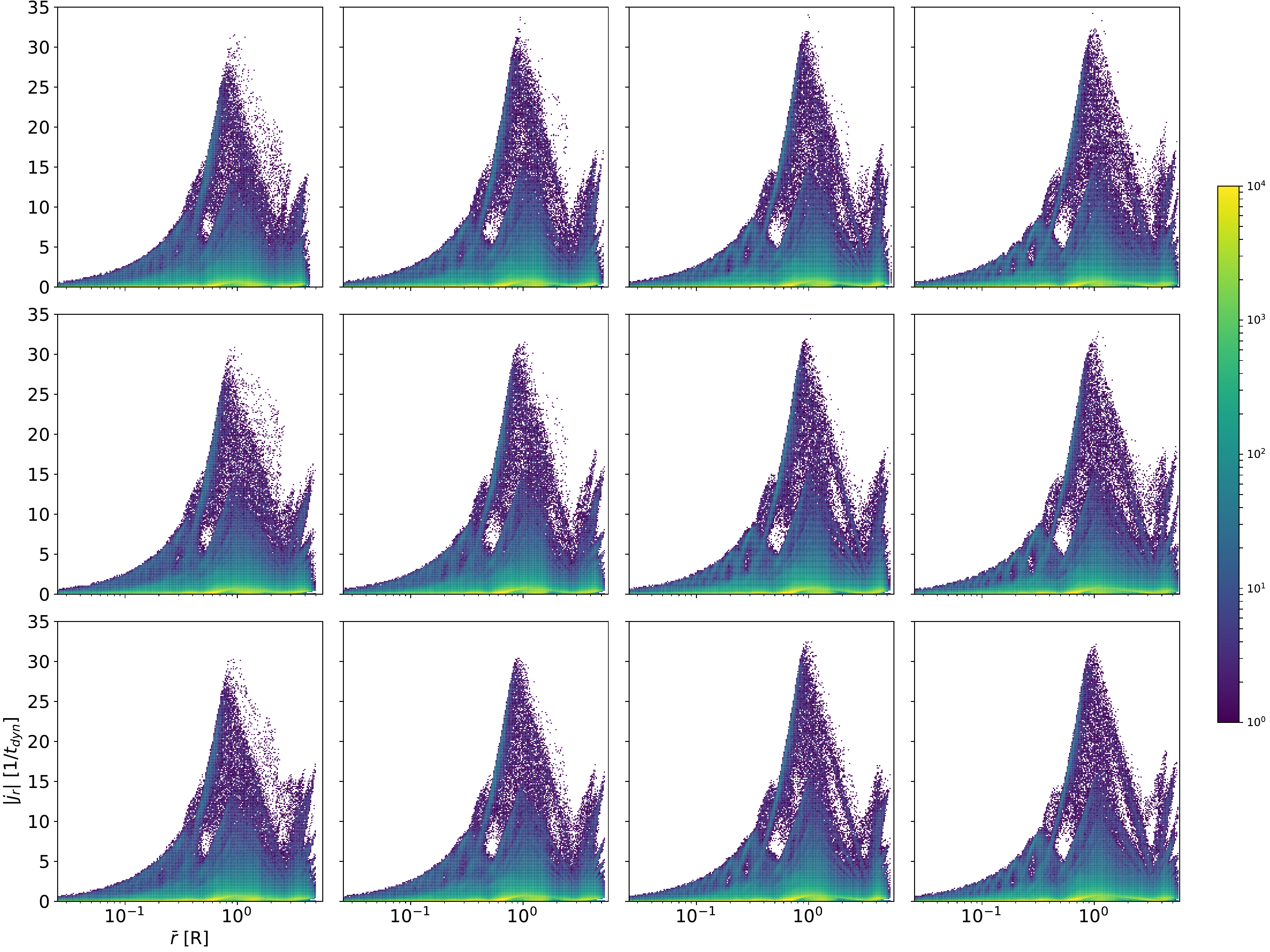}
\caption{Radial orbital action and mean orbit radius distributions of Gaussian isolated collapse simulations. panels are organized by correlation (column-wise: C=0.5,0.75,0.9,1.0 left to right) and spin (row-wise: $\lambda$=0.0,0.05,0.1 top to bottom). Samples are taken from $100$ equally-spaced frames during $10 - 12 t_{dyn}$. Sample points are taken such that each particle is given one mean radius and the radial auto-correlation power spectrum.}
\label{fig:rmeanraction}
\end{center}
\end{figure*}

The particle distribution in a self-gravitating virialized halo are known to be non-thermal, producing resonances and other attractors as a byproduct. Halo forces point primarily in the radial direction, and these strong forces are capable of setting the radial action into particular modes of motion. These resonances may appear as local fine structure in a halo's energy distribution or another features such as a turn-around caustic. Resonances are visible at all radii in all simulations, Fig.~\ref{fig:rmeanraction}. New structure is visible among these resonances. For example, counting in from the most prominent branch, which peaks at $ \bar{r} \sim 1 R$, one sees 8-9 distinct resonances in CDM halos running up from $\bar{r} = 0.04 R$. Noise and force softening subdues resonances much below $\bar{r} = 0.04 R$. The number of resonances increases with correlation, reaching 14 for $C=0.5$. The more crowded Bose spectra also look more self-similar, meaning uniform in shape, than their CDM counterparts. The inner orbital resonances are unfortunately at the limit of what can be resolved at the current levels of softening and time steps, though a number of other resonant or coherent features can be seen at larger radii through the virial radius. More resolved Bose fine structure will have to wait for improved simulations.

\section{Discussion}
\label{Discussion}

We now seek to further classify the differences in structure observed between Bose and classical halos. Estimating the causes of new structure will be a continuing process throughout the axion structure formation series as simulations and techniques become more sophisticated.

The initial infall produces the expected result of increased rapidity to first crossing, as predicted by the example of in-falling spherical shells in LQR, and expanded upon in Appx.~\ref{SShell}. Gaussian halos show a quickening of infall on the order of $O(30 \sqrt{1-C}\%)$. Top-hat halos widen that margin. The earlier infall would shift slightly the formation of early galaxies and other large structures to earlier cosmological times. Observational evidence and numerical models of early halo formation, through the epoch of reionization for example, currently do not favor such an outcome \citep{Mesinger2016}. The shift is only slight, fortunately, and not likely to make much of an impact on the re-ionization redshift. Faster crossing times and other properties of the XC force also leads to shorter times to virialization. Rate of equilibration among Gaussian Bose halos are seen to increase relative to classical halos by $\sim 70 (1-C) \%$.

Fluid elements of Bose DM are subject to forces larger than that of Newtonian gravity at the halo center, and weaker than at the halo's edge. Variance in force augmentation in Gaussian halos lines up well with velocity dispersion augmentation of those same halos. In the presence of near-identical spatial distributions, we hold the variable force augmentation as evidence of effective velocity dependence in the XC interaction. Further, there may be an opportunity to observe this force augmentation via the substructure of major halos. If it is found that halos and their substructure remain in correlated contact, it is possible that the experienced force of the Bose DM substructure is also altered relative to the classical case. The gravity felt by baryons within the dwarf galaxy would be dominated by the Newtonian potential and, being bound to the sub-halo, would be pulled along the modified orbit. The possible implications for substructure orbits and their organization will be pursued with great interest in future cosmological volume simulations.

\begin{figure}
\begin{center}
\includegraphics[width=8cm]{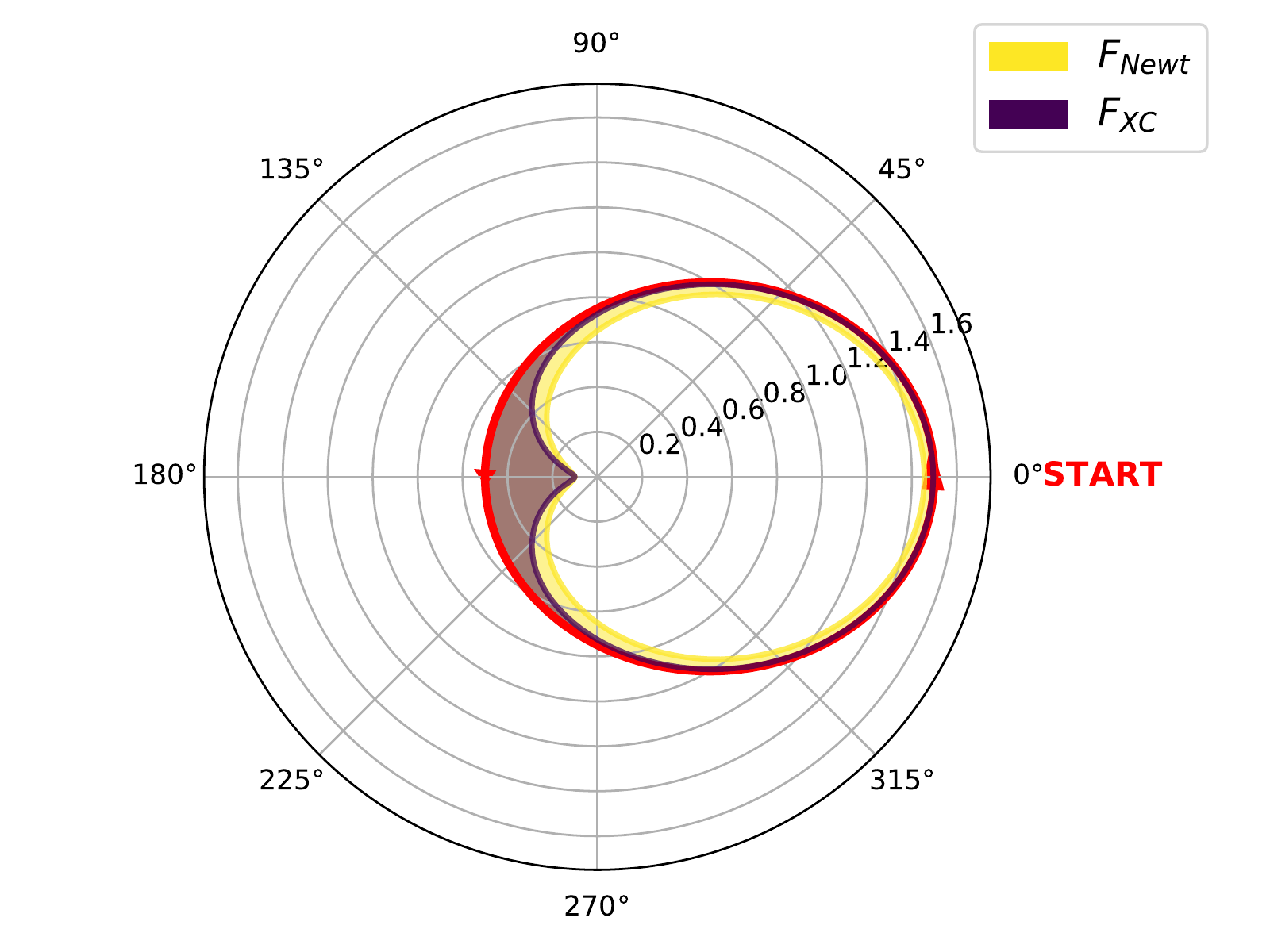}
\caption{Illustration of the force experienced by a test DM particle in a spherically symmetric Bose halo with density profile similar to that in the Top-hat simulations. The represented angle is the one canonical to the radial action. The forces extend radially from the orbit point toward the center of the plot. The classical Newtonian force (yellow) is periodic with maximum at the point of closest approach due to the gravitational potential depending only on the mass density distribution. The XC force (purple) is seen to vary more widely with radius, as the average phase-space density increases towards the center of the halo. Note that this orbit would not be closed in general.}
\label{fig:orbitdiag}
\end{center}
\end{figure}

The matching of virialized density and anisotropy profiles among Bose and classical halos is unexpected. The demonstrated difference in force and dependence of XC on the full phase-space density provide multiple opportunities for a halo to settle into a differently shaped state. Put another way, there appear to be mechanisms that one would expect to force the halo into a new configuration. For example, the XC force experienced by particles on highly radial trajectories is significantly different than the Newtonian force, Fig.~\ref{fig:orbitdiag}, implying that each halo would have very different orbital structures. The orbital analysis of the previous section did show a sizable change in both the number and size of orbital resonances within the virial volume. Though, again, the influence of these structures appears to be limited to the kinematical domain as there are no observed large scale migrations of halo halo spatial structure from the universal profile. A more precise investigation of a Bose halo's orbital properties, and how these properties play into reproducing the emergent scalse of classical halos, will be left for future work. It is also worth noting that constraint forces perform no work, making the XC's ability to transport energy through the distribution limited. Appendix~\ref{SShell} provides some further insight on the altered force profile via the examples of spherical shell collapse.

The role of the breaking radius $r_{break} = 0.4R$ in the Gaussian density profile is itself an interesting topic. The breaking radius is found to be insensitive to both spin and correlation. Many kinematical and dynamical transitions between classical and Bose halos also occur at this radius, such as the relative size of mean phase-space density, spherical velocity dispersion, force density, and energy density. This conspiracy of emergent scales may be indicative of significant physical phenomena important to both classical and Bose halos. This topic will be studied further in future work.

Changes to the classical energy profile can accumulate differences to the order of $O(400(1-C)^2 \%)$ in Gaussian collapse, and larger differences are observed in Top-hat halos. Unfortunately, current baryonic means of measuring a halo's energy density are unable to differentiate ab initio between Bose and classical structures. A more direct measure of the local DM energy distribution is needed to distinguish bulk correlated and uncorrelated motion. Remote detection efforts searching for axion decay may be capable of detecting such differences \citep{Kelley2017,Bull2018}.

Changes to structure, even at the present level of coarse graining, present windows to detection of the axion. The observed kinematical signatures of Bose physics such as augmented circular orbit speeds, enhanced velocity dispersion, and migration of the fractional speed distribution are all theoretically visible in direct and indirect axion detection efforts \citep{lentz2017,Foster2018,Bull2018,Knirck2018}. Further, violation of the classical binding energy condition among halo bodies is one of the more unique findings from Bose physics. Though the sub-population is small, on the order of a tenth of a percent for $C=0.5$, it would be a tell-tale signature of XC physics if detected. The absence of such a sub-population is not necessarily evidence of trivial XC physics though, as the sub-population's size also depends strongly on the shape of the initial distribution.

The compactness of orbital actions among Guassian halos leads one in the direction of interpreting XC as a force that maintains coherence, though the observables that are being protected have yet to be identified. The drive to cohere also appears to extend to the smoothing of classical boundaries such as the virial radius. Mixing across the virial barrier is very weak in classical infall. The XC force of Bose infall is known from Fig.~\ref{fig:circvr} to point radially outward at the virial radius. This outward-pointing effective force is suspected to effectively displace virialized orbits from their classical trajectories into the classically un-virialized space, smoothing the transition. The mixing can also be phrased in more traditional condensed matter framework. Standard Fermi systems near condensation have an analogous barrier, the Fermi surface. Modification of chemical potentials, including through XC interactions, often degrade the Fermi surface in the form of a theory of quasi-particles on the barrier \citep{Nozieres1964}. Bose systems at zero temperature are also capable of forming quasi-particles \citep{Leggett2006}. The non-local XC force of self-gravitating axions is still akin to a chemical potential, being largely set by Lagrange multipliers of particle number conservation. It is suspected that the XC's degradation of the virial barrier may be understood as the mixing of quasi-particle excitations. The role this mixing plays on the inter-galactic media and the medium between halo and substructure remains to be seen. It may prove that the distinction between different virial volumes becomes degraded, which may impact the shape, mass, and motions of cosmological halos and their substructure.

\begin{figure*}
\begin{center}
\includegraphics[width=\textwidth]{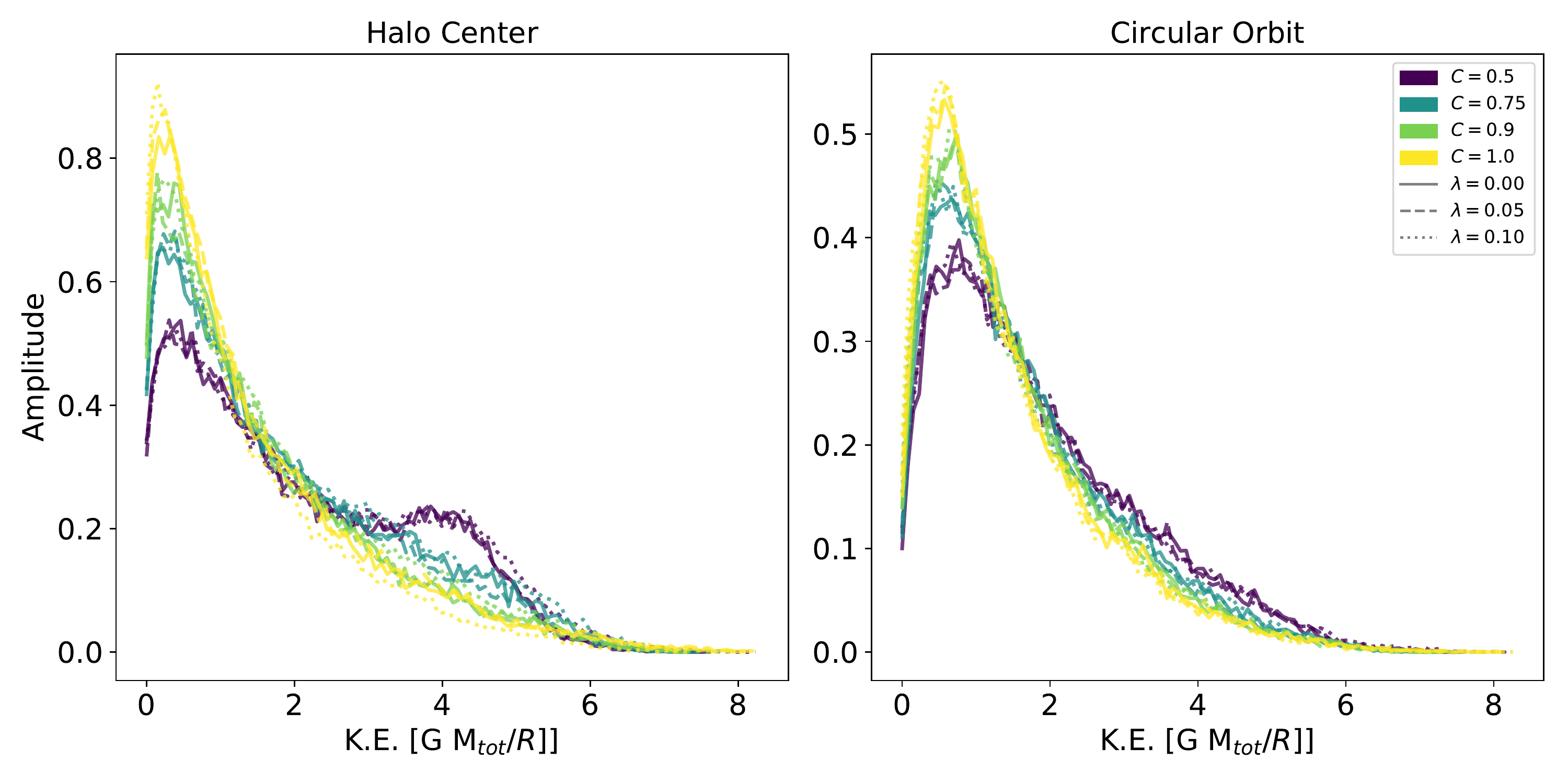}
\caption{Kinetic energy spectra of Gaussian isolated collapse simulations. Samples are taken from stacked frames of simulations of post-virialized halos, explained in detail in Section~\ref{finestruct}. The sampling region within each halo is chosen similarly to \citet{lentz2017}. Samples are taken from a toroid of extent $0.04 R\le s \le 0.06 R$ and $-0.01 R \le z \le 0.01 R$ in cylindrical coordinates. The origin of the coordinate system is placed at the halo center, and the orientation is such that $\hat{\bmath{z}}$ is parallel with $\bmath{j}_{\phi}$. Kinetic energies are measured either from a halo-centered frame of reference (left), or a frame on circular orbit at radius $s=0.05R$ that is co-moving with the halo's net spin (right). The co-moving frame is designed to emulate the Sun's motion in the Milky Way halo.}
\label{fig:signalshape}
\end{center}
\end{figure*}

The presented impacts of Bose physics on a halo's angular momentum and speed distributions have a profound effect on the local energy distribution. Axion DM searches using the decay of halo axions into photons are sensitive to these changes. We construct samples of local energy spectra at the approximate solar radius $r=0.05 R$, one fiftieth the halos' virial radius, to estimate the signal shapes as may be seen by various DM searches, Fig.~\ref{fig:signalshape}. The set of signal in the halo center frame, as may be observed in a radio telescope survey of a nearby galaxy \citep{Kelley2017,Bull2018}, show Bose structures that significantly depart from CDM and standard near-isothermal shapes. Depletion of the lone thermal-like peak in Bose spectra is accompanied by a large secondary peak at $K.E. =4 GM/R$. Either of these features may stand out as markers of Bose physics. Also, like the angular momentum and speed distributions of Figs.~\ref{fig:jdistr}, \ref{fig:vfracdistr}, the overall width of the Bose spectra are somewhat smaller than their CDM counterparts. The circularly-orbiting signals match what a terrestrial direct axion DM search may see \citep{MADMAX2017,Jeong2018,Du2018,Zhong2018}. The differences in shape seen in the halo frame signals are somewhat marginalized by boosting to a co-rotating circular orbit, though the new high energy structure is still clearly visible. Fine structure within the new high-energy peak, or elsewhere in the spectra, cannot be resolved using the current simulations. More resolved and complete simulations will be needed to provide realistic signal shapes with fine structure.

\section{Summary}
\label{Summary}

This paper extends the analysis of the highly-correlated QCD axion dark matter of \citet{Lentz2018b} into the non-linear regime of structure formation. The Boltzmann-like equation derived in \citet{Lentz2018b} to govern the dynamics of a degenerate correlated Bose fluid is first re-introduced, and the exchange-correlation interaction is reviewed. The numerical algorithms for an N-Body method are obtained from the continuum equation of motion by using the method of characteristics, discretizing the continuum distribution function into integral particles whose individual equations of motion can be different from standard Newtonian gravity. The position and velocity dependence of the exchange-correlation force are discussed.

Simulations of spherical and rotated collapse are run in a static cosmology using the small-scale N-Body solver, Condensate in eXternal Potential. Comparisons between Bose physics and the standard pressure-less cold dark matter model consistently show unique Bose structures during initial infall, chaotic relaxation, and into virialization. Novel Bose structures are found to include: 
\begin{itemize}

    \item Significantly augmented forces on dark matter fluid elements

    \item Increased rapidity of first infall by $\gtrsim 30 \sqrt{1-C} \%$ and faster rates of equilibration by $\gtrsim 70(1-C)\%$
    
    \item Shifts in the classical binding energy profile to the order of $\gtrsim 400(1-C)^2\%$
    
    \item Altered orbital characteristics of DM, including increases and decreases in circular orbit speed by $\gtrsim 130 (1-C)\%$
    
    \item New substructure among classical energy and angular momentum distributions
    
    \item Small sub-populations of halo particles that violate the classical binding condition of virialized systems
    
    \item New macroscopic populations at moderate energy or speed relative to classical limits
    
    \item Concentration of kinetic distributions about new substructure
    
    \item Concentration of orbital actions in all three independent dimensions
    
    \item New and increasingly self-similar orbital resonances and other fine structures
    
    \item Evidence of correlation-induced mixing across the virial radius

    \item Alignment of the universal emergent scale in the halo mass density profile with Bose kinematical and dynamical transition scales 
    
    \item Notable \textit{lack} of impact to the halo density and anisotropy profiles
\end{itemize}
The above novel structures already suggest several avenues for search among direct and indirect axion DM experiments, and in the observation of baryonic processes.

The presence of surviving Bose-specific structures in simulation is a significant development in the search for the dark matter. The presented small-scale simulations are far from sufficient to classify all Bose structure, however. Higher resolution, the incorporation of baryons, and the use of cosmologically-motivated initial conditions are needed to better tabulate the structural differences of a Bose fluid in a realistic and natural environment. The next installment of this series builds on the small-scale tests of the generalized Bose dynamics, incorporating them into the highly sophisticated N-Body+Smoothed-Particle-Hydrodynamics code ChaNGa \citep{Menon2015}. The ability to increase resolution and fold in well-modeled non-dark matter species in a cosmological setting will provide the first glimpse into the realistic extent of unique axion structures. 

Lastly, there are two fundamental challenges to address, in order in order to create a fully physically-motivated model of axion structure formation. First, adding baryonic species to a system of highly degenerate bosons influences not only the classical gravitational potential of that system, but may also alter the completeness of the condensate.  The derivations of ASF1 are made in the environment of minimal influence from baryonic gravity. This condition may not necessarily be satisfied in the centers of galaxies, where baryonic densities greatly exceed those expected of DM. Specifically, the addition of significant gravitating species `external' to the Bose fluid disrupts the symmetry used to solve the many-body Schr\"odinger equation of \citet{Lentz2018b}, a symmetry which permitted condensate solutions to be set as products of two-body correlators. External potentials can therefore introduce a migration of states in the highly-degenerate fluid, meaning the complete condensate formalism may not hold in the cosmological context. The effective size of those influences on the correlation is a topic of ongoing research.

Secondly, it is not obvious is how much correlation exists within these systems of axions at the start of the collapse simulations. Initial correlation has so far been a tune-able integral parameter. A great deal of physics and different dynamical regimes occur between the parent scalar field and post-recombination structure formation. A complete description of state dynamics within and outside of the degenerate state of DM axions, from the well-motivated conditions of the pre-inflation era to the matter era, is also a topic of current study. A dynamical description of state tracking will further the goal of resolving the extent to which Bose physics can create structure unique from standard cold dark matter, and is planned to be introduced in subsequent papers.

\section{Acknowledgments}
\label{Acknowledgments} 

We would like to thank Jens Niemeyer, Katy Clough, David Marsh, Bodo Schwabe, Jan Velmatt, Xiaolong Du, Marcel Pawlowski, and Ewald Puchwein for their productive discussions in the refinement of this work. We also gratefully acknowledge the support of the U.S. Department of Energy office of High Energy Physics and the National Science Foundation. TQ was supported in part by the NSF grant AST-1514868. EL and LR were supported in part by the DOE grant DE-SC0011665.

\appendix

\section{Method of Characteristics and leapfrog Integration}
\label{MOC}

The Boltzmann-like system of axion infall
\begin{align}
    &0 = \partial_t f^{(1)} + \frac{\bmath{v}}{a^2} \cdot \bmath{\nabla} f^{(1)} -  \bmath{\nabla} \bar{\Phi} \cdot \bmath{\nabla}_v f^{(1)} \nonumber \\
&- m_a \bmath{\nabla}_v \cdot \Bigg( f^{(1)} \bmath{\nabla} \int d^6 w_2 f^{(1)}(w_2,t) \phi_{12} \nonumber \\
& \times \left( \frac{C-1 - \left(\lambda_1+\lambda_2\right)f_+}{1+ \lambda_2 f_+} \right)\Bigg) \label{BoseEOM}
\end{align}
falls into the hyperbolic class of partial differential equations (PDEs). This class, which includes conservation laws and wave equations, is amenable to many elegant numerical solvers including the method of characteristics (MOC). A Lagrangian method, the method of characteristics propagates integrals of motion to a system of differential equations through parameter space, forming an accurate sample of the solution. For an extensive review of the technique, see \citet{Courant1953}. Conveniently, the Bose system equation of motion is of first order in derivatives, requiring only a rudimentary level of understanding of MOC. Time integration using the leapfrog method is also presented.

\subsection{Method of Characteristics}

\subsubsection{Theory}

Using the notation of \citet{Courant1953}, the Boltzmann-like equation of motion of the correlated Bose fluid falls into the class of PDEs of the form
\begin{equation}
F(\{x_i\},u,\{p_i\}) = 0 \label{MOCF}
\end{equation}
where $\{x_i\}$ are the $n$ coordinates, $u$ is the function to be solved for, dependent on the $x_i$, $\{p_i\}$ are those fluxes given by partial derivatives $p_i = \partial_{x_i} u $, and the form of function $F$ is first-order smooth in its arguments. $F$ is seen to be an integral of motion. If $u$ is any solution to the defining equation ($F=0$), let us construct a curve in $\Re^{2n+1}$ of $\left( \{x_i(s)\}, u(s), \{p_i(s)\}\right)$ such that $u(s) = u(x_i(s))$. Differentiating Eqn.~\ref{MOCF} along the curve gives 
\begin{equation}
\sum_i \left( \partial_{x_i} F + \partial_u F p_i\right) \dot{x}_i + \sum_i \partial_{p_i} F \dot{p}_i = 0
\end{equation}
A constraint equation can also be found in $u$ from the curve speed
\begin{equation}
\dot{u} - \sum_i p_i \dot{x}_i = 0
\end{equation}
More generally, the implicitly differentiated form $du - \sum_i p_i dx_i = 0$ also holds. A second constraint may be found from using the exterior derivative language.
\begin{align}
0 &= d\left(du - \sum_i p_i dx_i\right) \nonumber \\
&= \sum_i \left(dp_i \dot{x}_i - \dot{p}_i d x_i \right) 
\end{align}
Altogether these relations may be organized to provide equations of motion for the solution along the path
\begin{align}
\dot{u} &= \sum_i p_i F_{p_i}, \\
\dot{x}_i &= F_{p_i}, \\
\dot{p}_i &= - F_{x_i} - F_{u} p_i.
\end{align}

\subsubsection{Application}

For our application to the first-order axion Boltzmann-like equation, we map $u \to f$, $\{x_i\} \to (\bmath{x},\bmath{v})$, $\{p_i\} \to (\bmath{\nabla} f,\bmath{\nabla}_v f)$, and $F$ is the right-hand-side of Eqn.~\ref{BoseEOM}. We see again that $f$ is an integral of characteristic motion. Further the equation form appears to be much like that from a Hamilton's principle with Hamiltonian $F$. We choose time to be the suitable parameterization of the characteristic. The characteristic equations are specifically found to be
\begin{align} 
\dot{f}_1 &= 0 \\
\dot{\bmath{x}} &= \frac{\bmath{v}}{a^2 } \\
\dot{\bmath{v}} &= -\bmath{\nabla} \bar{\Phi} \nonumber \\
& - m_a \frac{\partial}{\partial \bmath{\nabla}_v f_1} \int d^6w_2 \bmath{\nabla} \phi_{12} \cdot \bmath{\nabla}_v \left(f_1 \frac{C-1 - \lambda_+ f_+}{1 + \lambda_2 f_+} f_2\right),
\end{align}
where the dot total derivatives are now with respect to time, and $w_1 = (\bmath{x},\bmath{v})$. For completeness, the functional derivative evaluates to
\begin{align}
    &\frac{\partial}{\partial \bmath{\nabla}_v f_1} \int d^6w_2 \bmath{\nabla} \phi_{12} \cdot \bmath{\nabla}_v \left(f_1 \frac{C-1 - \lambda_+ f_+}{1 + \lambda_2 f_+} f_2\right) = \nonumber \\
    & \int d^6w_2 \bmath{\nabla} \phi_{12}  \left( \frac{C-1 - \lambda_+ f_+}{1 + \lambda_2 f_+} f_2\right) \nonumber \\
    &+ \int d^6w_2 \bmath{\nabla} \phi_{12}  \left(f_1 \frac{- \lambda_+/2}{1 + \lambda_2 f_+} f_2\right) \nonumber \\
    &- \int d^6w_2 \bmath{\nabla} \phi_{12}  \left(f_1 \frac{\lambda_2/2 (C-1 - \lambda_+ f_+)}{(1 + \lambda_2 f_+)^2} f_2\right).
\end{align}
We can clearly see how the physics of individual sample points differ from the standard Newton's second law of classical CDM literature, sourced by the non-trivial correlation function. It is worth noting that the solution to these samples are exact to the order of accuracy of the Boltzmann equation, so long as the potentials and forces are known. The parabolic Poisson PDE for the gravitational potential must also be solved at each distribution sample, in order to calculate the gravitational force contribution. Techniques for solving the gravitational potential of an N-Body sample often implement some artificial smoothing length to represent the contribution from a representative volume of phase space about a sample point; this  is where exactness of modern gravitational MOC N-Body implementations breaks down, though it is still very productive.

In practice, MOC is used to track many sample characteristics from some initial Cauchy surface to a prescribed end time. As the DF is constant along these characteristic curves, choosing sample points according to distribution weight creates an effective distribution sampling from which equal-time observables may be calculated. These sample points are often referred to as particles or bodies, though they do not map directly to individual axions. Sampling of the DF often looks like an optimization between accurately reproducing the interactions and providing a clear means of interpreting the output configuration. In terms of the chosen gravitational sampling kernel of Section~\ref{CXP}, the total force felt by an N-body particle amounts to
\begin{align}
    &\bmath{F}_1 = -\sum_i^n \bmath{\nabla} \Phi_{1i}\nonumber \\
    &- \sum_i^n \bmath{\nabla} \Phi_{1i}  \left( \frac{C-1 - \lambda_+ \left( f_1 + f_i\right)/2}{1 + \lambda_2  \left( f_1 + f_i\right)/2}\right) \nonumber \\
    &- \sum_i^n \bmath{\nabla} \Phi_{1i}  \left( \frac{- \lambda_+ f_1 /2}{1 + \lambda_2  \left( f_1 + f_i\right)/2} \right) \nonumber \\
    &+ \sum_i^n \bmath{\nabla} \Phi_{1i}  \left(f_1 \frac{\lambda_2/2 (C-1 - \lambda_+  \left( f_1 + f_i\right)/2)}{\left(1 + \lambda_2  \left( f_1 + f_i\right)/2 \right)^2} \right), \label{XCforceappx}
\end{align}
where $\Phi_{ij}$ is the inter-sample K1 Newtonian potential kernel.

\subsection{Leapfrog Symplectic Integration with Exchange-Correlation Force}

After the method of characteristics, there is still one dimension left to partition before the axion fluid can be translated to machine language. Both the continuum and phase space discretized sample points derived above preserve the structure of phase space via $\dot{f}=0$ and integral masses. We choose the leapfrog symplectic integrator to map the distribution over successive time steps as it is also designed to preserve volumes in phase space.

From the MOC derivations above, we can see that the particle equations of motion allow for a first order operator equation interpretation
\begin{equation}
\dot{w} = \hat{O} w,
\end{equation} 
where $w = (\bmath{x}, \bmath{v})$ and $\hat{O}$ is the system's evolution operator. The evolution operator of the sample can be written as
\begin{equation}
    \hat{O} = \hat{T} + \hat{V}_{\bar{\Phi}} + \hat{V}_{XC},
\end{equation}
where $\hat{T}$ is the standard non-relativistic kinetic operator, $\hat{V}_{\bar{\Phi}}$ is the mean field gravity operator, and $\hat{V}_{XC}$ is the XC operator. The leapfrog algorithm requires that the evolution operator is separable over its position and momentum dependence. Fortunately each sample's distribution function value is invariant along characteristic curves, removing explicit velocity dependence in the XC force, at least the order of accuracy of the integration scheme. It is therefore possible to separate the operator equation of motion into components exclusively dependent on the velocity or position sub-spaces of phase space. Leapfrog integration can therefore be derived in the usual way. Applied to our axion system, the equations of motion for a sample particle become
\begin{align}
\bmath{x}(t/2) &= \bmath{x}(0) + \frac{\bmath{v}(0)}{a^2 } \frac{t}{2}, \\
\bmath{v}(t) &= \bmath{v}(0) -t \Bigg( -\sum_i^n \bmath{\nabla} \Phi^{K1}_{1i}\nonumber \\
    &- \sum_i^n \bmath{\nabla} \Phi^{K1}_{1i}  \left( \frac{C-1 - \lambda_+ \left( f_1 + f_i\right)/2}{1 + \lambda_2  \left( f_1 + f_i\right)/2}\right) \nonumber \\
    &- \sum_i^n \bmath{\nabla} \Phi^{K1}_{1i}  \left( \frac{- \lambda_+ f_1 /2}{1 + \lambda_2  \left( f_1 + f_i\right)/2} \right) \nonumber \\
    &+ \sum_i^n \bmath{\nabla} \Phi^{K1}_{1i}  \left(f_- \frac{\lambda_2/2 (C-1 - \lambda_+  \left( f_1 + f_i\right)/2)}{\left(1 + \lambda_2  \left( f_1 + f_i\right)/2 \right)^2} \right) \Bigg)\Bigg|_{\bmath{x}(t/2)}, \\ 
\bmath{x}(t) &= \bmath{x}(t/2) + \frac{\bmath{v}(t)}{a^2 m} \frac{t}{2}.
\end{align}
where again the subscript "$1$" indicates evaluation at $(\bmath{x},\bmath{v})$.

\section{Intuition for Axion Collapse with Spherical Shells}
\label{SShell}

Reducing the dimension of the full phase-space problem of Eqns.~\ref{Boltz}, \ref{Poisson}, \ref{lc} may help to build an intuitive understanding of the physics involved. This subsection explores the collapse of angularly-rigid spherical shells in both the static and rotational cases.

\subsection{Spherical Shell}

\begin{figure} 
\begin{center}
\includegraphics[width=9cm]{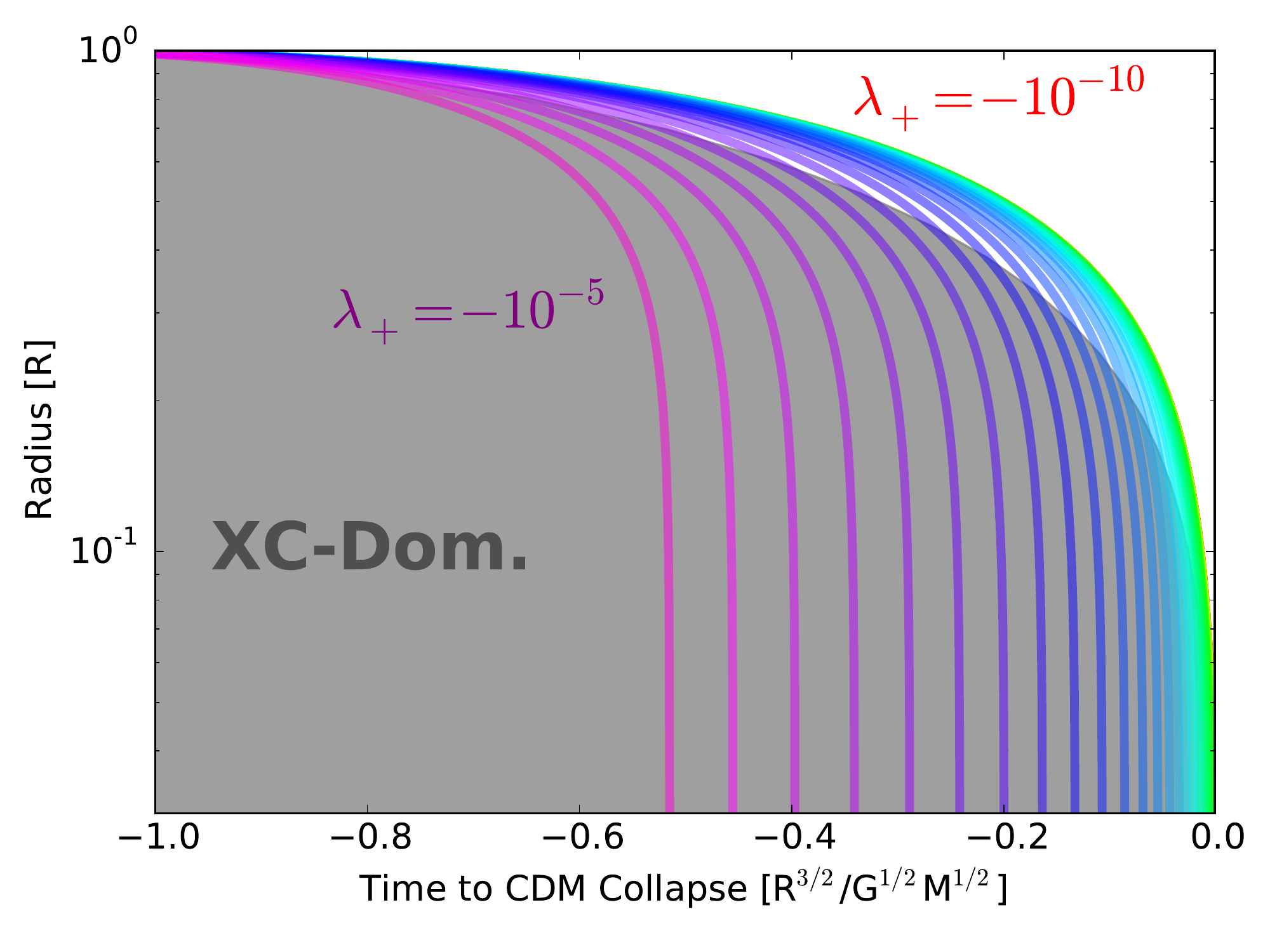}
\caption{Collapse time to radius of spherical collapse from rest. The color gradient is logarithmic in $\lambda_+$, running from $\lambda_+=-10^{-10}$ at the bottom to $\lambda_+=-10^{-3}$ at the top. The region where post-Newtonian physics dominates over classical gravity is shaded in gray. }
\label{sph_shell_coll}
\end{center}
\end{figure}

This example is largely similar to the demonstration of \citet{Lentz2018a}. A thin and cold spherical shell may collapse under its own gravity. For simplicity, we have additionally assumed no tangential axion motion, that the radial velocity dispersion is sufficiently small so as to leave the shell width unchanged over the simulated collapse. These symmetries reduce the dimensionality of the problem from $7$ to $1$. We also use constraints of $C=1$, and $|\lambda_2 f_+| \ll 1$ to simplify the contribution of exchange-correlation interactions. Finally, $\lambda_+$ is taken as a free tunable parameter for the purpose of demonstrating the range of impact from XC interactions. The governing equation of the Bose collapse may then be written to leading order in dispersion as
\begin{align}
\ddot{r} &= -\frac{GM}{r_{soft}^2} \nonumber \\
&+\lambda_+\left(\frac{3 G M N(r_{soft})}{4 r_{soft}^2} - \frac{G M N'(r_{soft})}{4 r_{soft}}\right),
\end{align}
where $M$ is the effective gravitating mass, $N(r_{soft})$ is defined by
\begin{equation}
N(r_{soft}) = \frac{1}{8 \pi^2 r_{soft}^2 \sigma_r \sigma_v}
\end{equation}
where $r_{soft}^2 = \left(r^2 + \sigma_r^2\right)$, $\sigma_r$ is the width of the shell, $\sigma_v$ is the velocity dispersion of the shell and is held sufficiently small as to leave the shell width unchanged over the collapse. Note that the effective repulsion of the second two terms in the post-Newtonian contribution will never dominate over the first. Solutions of this system show that infall is amplified for non-trivial post-Newtonian contributions, Fig. \ref{sph_shell_coll}. The new physics leads to a characteristically more violent infall due to the sharper form of the $\sim 1/r_{soft}^4$ central XC force. Parameters chosen for spherical collapse are $r(t=0)=1$, $\sigma_r=10^{-3}$, and $\sigma_v=10^{-4}$ in dynamical units.

This simple example can be used to better understand the results of irrotational N-Body collapse. The classical homogeneous sphere is expected to collapse into a near-point under Newtonian gravity characteristic crossing time. The Bose homogeneous sphere is also expected to fall into a near-singularity as the XC force has the same dependence on effective gravitating mass. The Bose sphere is also expected to collapse more quickly and have greater speed at shell crossing than classical infall. In effect, the most violent collapse of classical self-gravitating systems is made more violent by exchange and correlation. Gaussian collapse is somewhat more difficult to interpret as the density and correlation function vary throughout the initial distribution, though the basic result of more rapid initial infall is still observed. The faster collapse may also provide an explanation for the increased velocity dispersion in the virialized halos.

\subsection{Rotating Spherical Shell}

Spinning the spherical shell about a central axis introduces the angular momentum pseudo-potential barrier to the system. The equation of motion for the shell is updated to 
\begin{align}
\ddot{r} &= -\frac{GM}{r_{soft}^2} + \frac{4 L^2}{3 \pi r_{soft}^3} \nonumber \\
&+\lambda_+\left(\frac{3 G M N(r_{soft})}{4 r_{soft}^2} - \frac{G M N'(r_{soft})}{4 r_{soft}}\right),
\end{align}
where $L = M \omega R^2$ is the angular momentum of the respective cylinder of radius $R$. Normalization and other factors are kept the same as in the non-rotating case. 

\begin{figure} 
\begin{center}
\includegraphics[width=9cm]{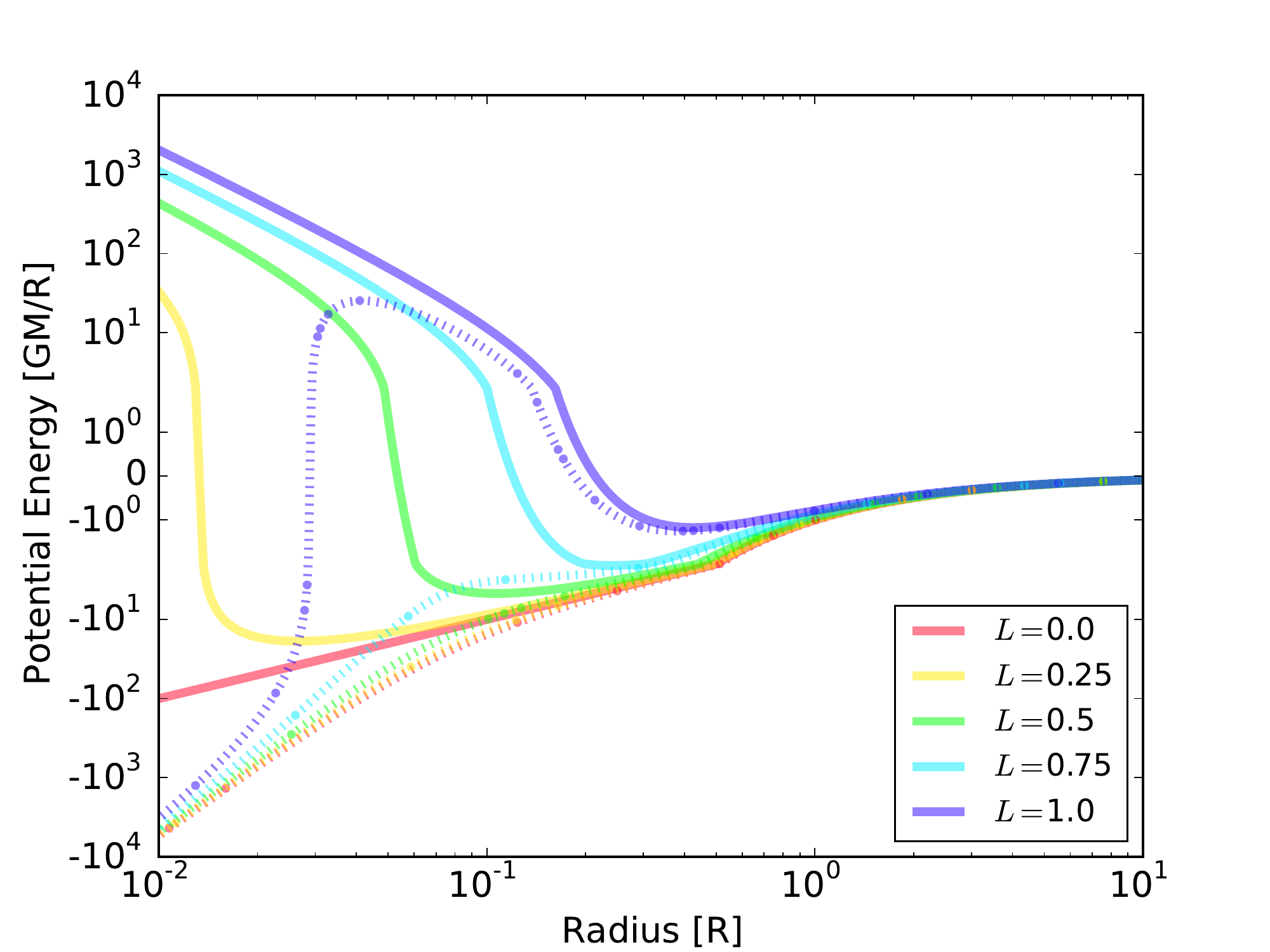}
\caption{Pseudo-potential of rotating spherical shell over radius. Solid lines correspond to classical physics, dashed lines to Bose physics at $\lambda_+=-10^{-7}$. Angular momentum of shell,  $L$, is quoted in units of $R^2 \omega M^{1/2}/G^{1/2}$. Kinks seen in lines are an artifact of the linear-to-log plotting scheme and are not features of the potentials.}
\label{rot_pot_coll}
\end{center}
\end{figure}

The presence of three singular forces allows for several possibilities in dynamics, Fig. \ref{rot_pot_coll}. The angular momentum pseudo-potential is singular, repulsive and of scaling $\sim 1/r_{soft}^2$, in contradiction to the attractive $\sim 1/r_{soft}$ classical gravitational attraction, and the $\sim 1/r_{soft}^3$ Bose attraction, producing an island between the classical-dominated and Bose-dominated regions.  The rotational island  produces a region of relatively slowed motion, but for small $L$ the angular momentum barrier is broken and collapse to singularity continues. An unstable stationary point exists when the barrier equals the shell's initial energy. Higher rotation re-establish the classical turn-around. As expected, there are accelerations in the rate and increases in the depth of Bose collapse.

\begin{figure} 
\begin{center}
\includegraphics[width=9cm]{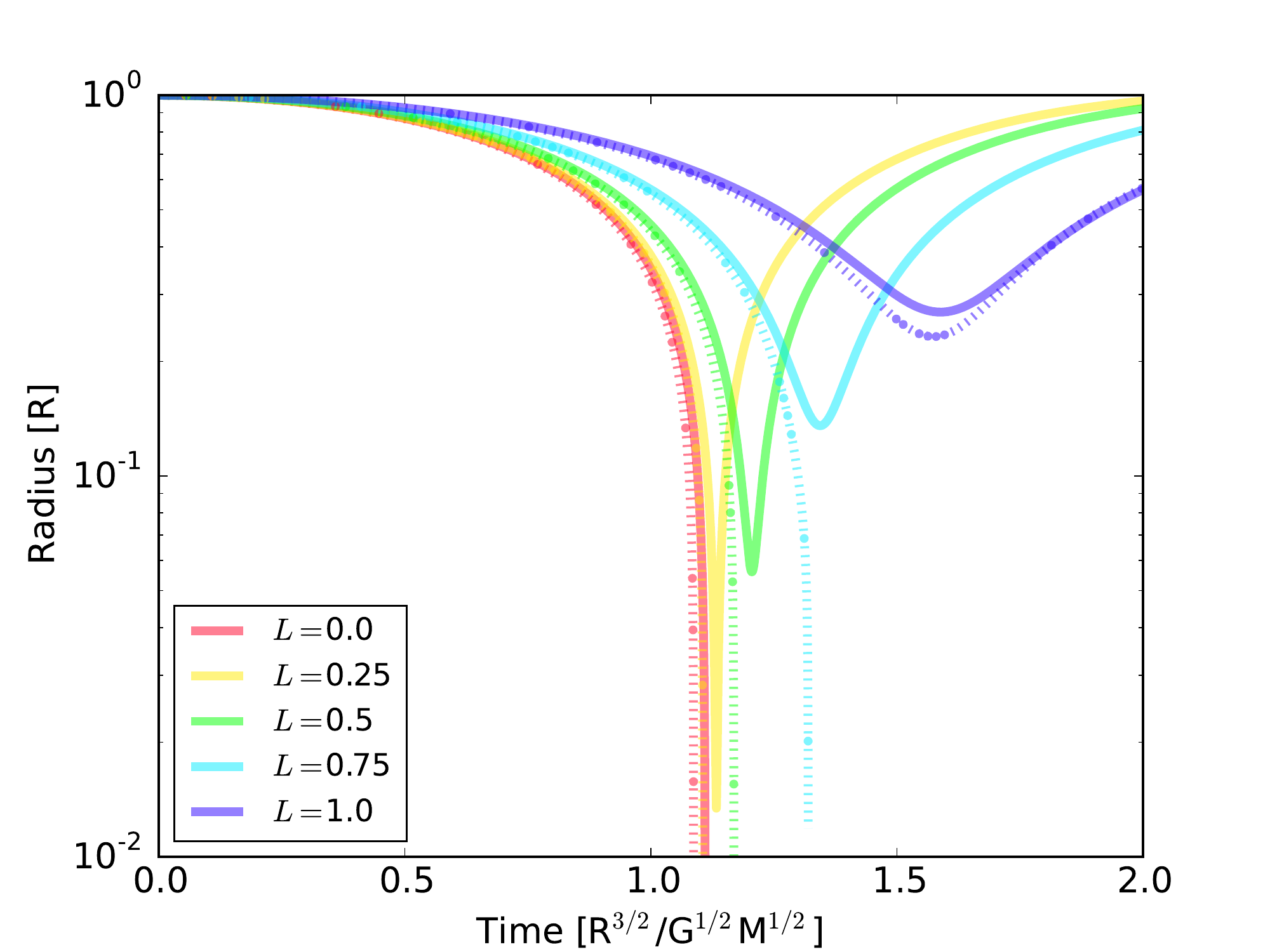}
\caption{Trajectories of rotating spherical shell collapse over time. Solid lines correspond to classical physics, dashed lines to Bose physics at $\lambda_+=-10^{-7}$. Angular momentum of shell,  $L$, is quoted in units of $R^2 \omega M^{1/2}/G^{1/2}$. Bose trajectories without rebound are physical as the Bose attractor has sufficiently degraded the angular momentum singularity.}
\label{rot_shell_coll}
\end{center}
\end{figure}

The spinning sphere simulations provide an explanation for the increased density of Bose Top-hat halos. The degradation of the angular momentum barrier by XC forces allows the spinning homogeneous sphere to collapse more completely than would otherwise be possible. The increased depth of the collapse then increases the central density of the virialized halo. Degradation of the angular momentum barrier may also provide some insight as to how difficult it may be to break the universal halo shape among Gaussian and more cosmologically relevant infalls.

\subsection{Lessons}

The primary lesson to be learned from these single-degree examples is that the increased scaling of the XC force over the Newtonian interaction in regions of high density. In the case of the infalling shell, an increasing phase-space density causes not only a faster rate of collapse, but a faster scaling of the new force. This increased scaling allows for the overtake of other pseudo-forces such as that from per-particle angular momentum conservation. The scaling of the XC force may impact the process of accretion and the capture of substructure in a halo, and will be a point of interest for cosmological simulations. 

Also note the role that momentum dispersion plays, namely how it enters the XC force similarly to spatial dispersion. If the velocity dispersion were allowed to evolve, it would increase under the acceleration of infall, and potentially diminish the acuteness of the force. The interaction of XC with the full phase space will be a point of interest in future publications.

\section{Structure Convergence}
\label{ConvStdy}

This study contains a number of limiting numerical elements, such as finite time steps, force softening, and sample number. We perform several tests on the simulation output to quantify and minimize the impacts of these non-physical parameters in the main analysis. The highlights of those analyses are provided here. We also look into the behavior of finer structures with respect to the parameter of particle number.

\begin{figure}
\begin{center}
\includegraphics[width=9cm]{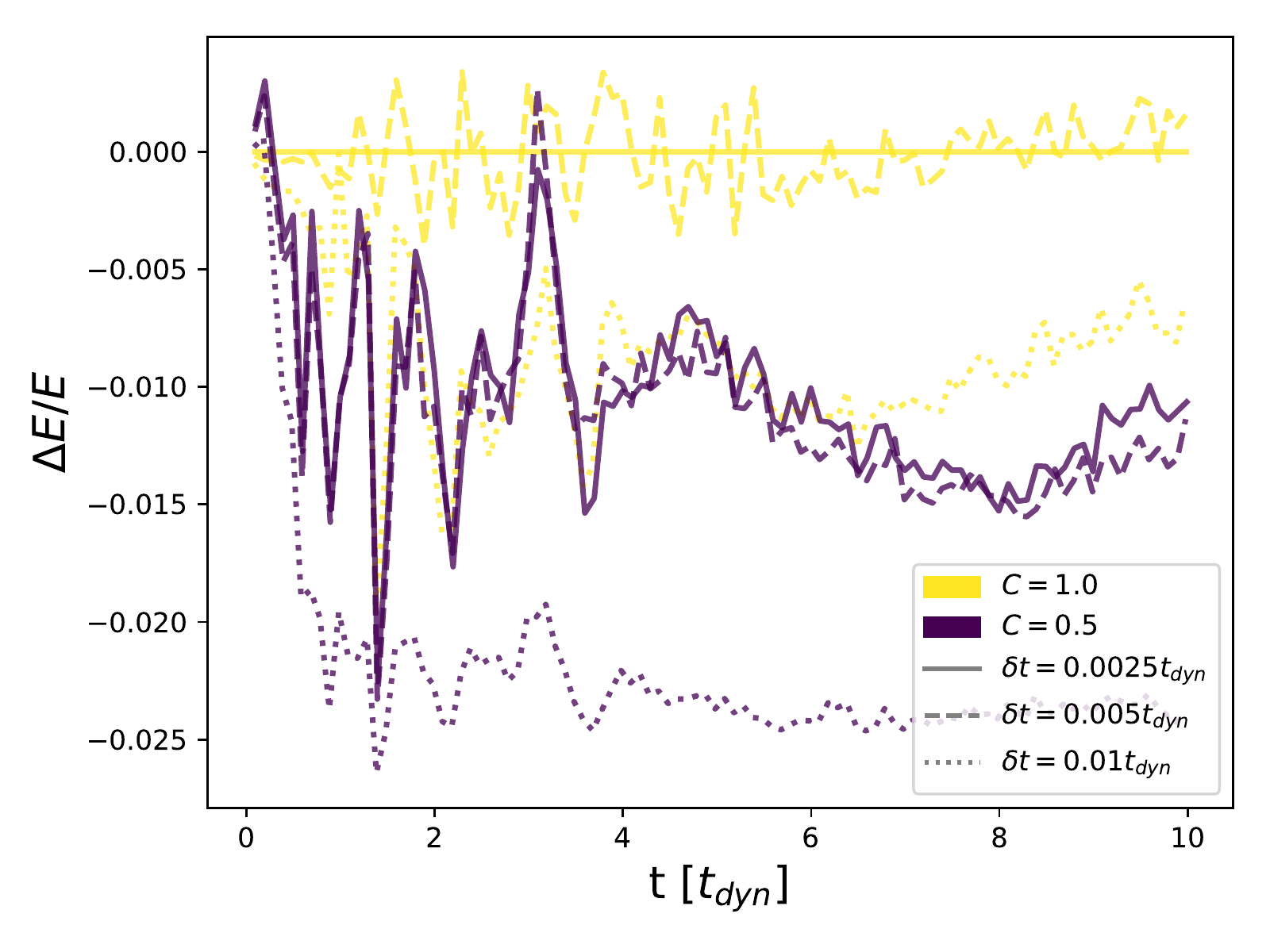}
\caption{Relative total classical energy of halo over time for Gaussian isolated collapse simulations of Sec.~\ref{SCS}. Energy is relative to classical collapse with time-step resolution $0.25\times10^{-2} t_{dyn}$. The period displayed is the $10 t_{dyn}$ of initial simulation. Profile coloration indicates degree of correlation ranging from classical $C=1.0$ to highly correlated $C=0.5$. The halos have no spin. Line style is given by step size in units of dynamical time.}
\label{fig:Econs}
\end{center}
\end{figure}

Finite integration of forces and kinetics over time produce minor errors in the evolution of degrees of freedom, and in the continuum integral actions of motion. These errors may accumulate over successive integrations, altering the resulting state, especially in regions of fast evolution. Leapfrog is specially chosen for its stability with respect to phase-space integrals. Total classical energy is an example of one such quantity, Fig.~\ref{fig:Econs}. Energy conservation is well converged in Gaussian infall for time-steps of $0.5 \times 10^{-2} t_{dyn}$ or smaller. Bose Gaussian halos do show an accumulated bias on the $1\%$ level relative to the $\delta t = 0.25 \times 10^{-2} t_{dyn}$ classical halo standard. The bias may be a consequence of XC and softening length interactions and will be investigated in future work. Top-hat halos behave similarly, though they require an increased softening length to minimize noise errors at the near-singular first crossing, and still produce slightly higher error around first crossing. Force evaluation and kinetic evaluation are performed at double precision to minimize floating point contributions to these errors.

\begin{figure*}
\begin{center}
\includegraphics[width=\textwidth]{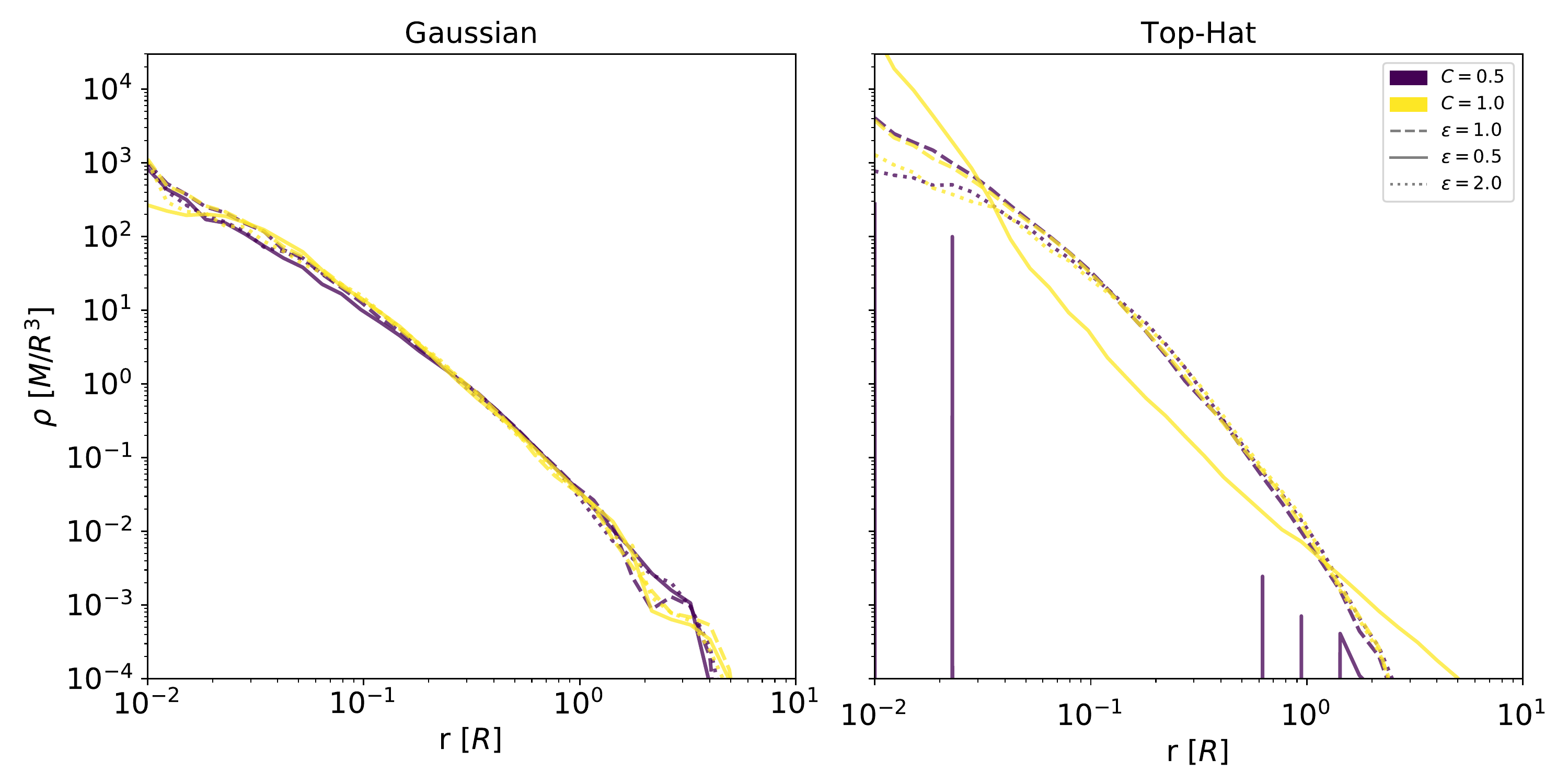}
\caption{Radial density profiles of the isolated collapse simulations of Sec.~\ref{SCS}. Gaussian profiles on L, Top-hat profiles on R. Halos were measure after evolving for $10 t_{dyn}$. Profile coloration indicates degree of correlation ranging from classical $C=1.0$ to highly correlated $C=0.5$. Halos have no spin. Line style indicates size of K1 softening length. $\varepsilon$ refers to the softening size relative to the shape-dependent central value in the text.}
\label{fig:rhorconv}
\end{center}
\end{figure*}

Force softening limits the degree to which one can distinguish dynamical structure to some fraction of the softening length. The general rule for the simulations presented in the main text is not to trust features much below the softening scale. Softening kernel shape or extent can impact the halo's global structure via shot noise fluctuations in the force from too large inter-particle separations, or forcing biases from too complete of overlap. There is an optimum length in between. One may find an optimal solution with direct statistical techniques such as in \citet{Dehnen2001}, though we use the more heuristic approach of observing robust profiles such as density, Fig.~\ref{fig:rhorconv}. For 50,000 particle Gaussian halos, smoothing much above or below $\epsilon = 1$, the sample-free softening length parameter of Eqn.~\ref{softlength}, produces an artificial depletion of mass at small radii, or other displacement of interior mass. Classical halos appear to be somewhat more susceptible than Bose. Top-hat collapses show an optimal scale at $\epsilon = 2$, much below which and the force noise disrupts the violent relaxation process. This sends the classical simulations nearer to a single power law halo. Bose collapses appear especially susceptible to softening length change, and scattering of many of the particles at singularity occurs if the softening is insufficient, resulting in a diffuse halo. Other aspects of XC sensitivity to softening are discussed below.

Total sample number provides the most direct subdivision of phase space as occupied by the presented halos. Sample number impacts the simulations' degree of resolution in multiple dimensions. An adjusted softening length abates the force errors some, but then introduces an enlarged intrinsic scale. The reduction in phase-space resolution leads to errors in observables of multiple forms.  We show these errors with respect to particle number for several figures of interest of the main text. The softening length changes with particle number according to Eqn.~\ref{softlength}.

\begin{figure}
\begin{center}
\includegraphics[width=\columnwidth]{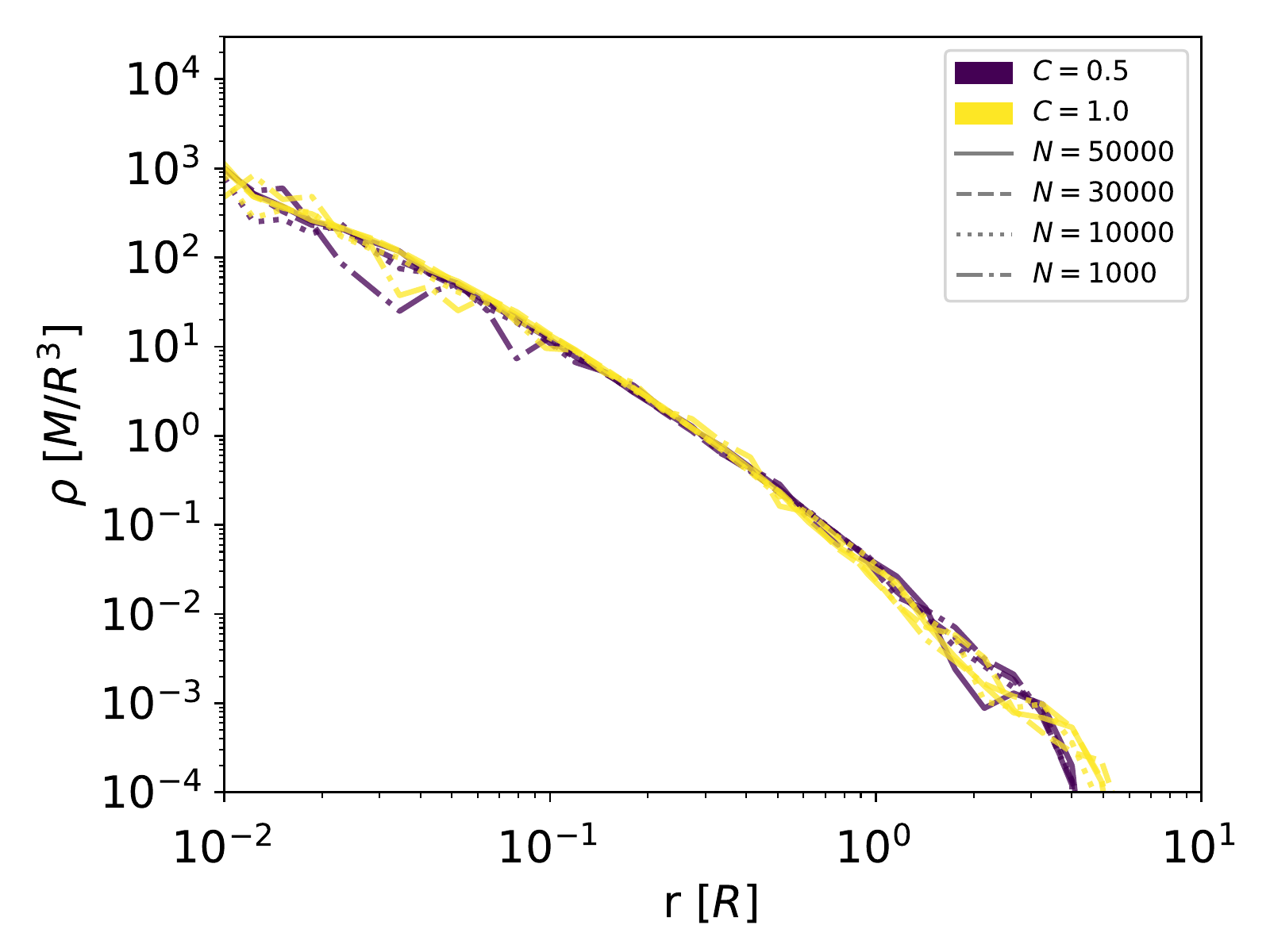}
\caption{Radial density profiles of the Gaussian isolated collapse simulations of Sec.~\ref{SCS}. Halos were measure after evolving for $10 t_{dyn}$. Halos are without spin. Profile coloration indicates degree of correlation ranging from classical $C=1.0$ to highly correlated $C=0.5$. Line style indicates number of particles. Simulations above 10,000 samples use softening parameter $\epsilon=1$, while 1,000 particle simulations use $\epsilon=2$ for the classical case and $\epsilon=3$ for Bose.}
\label{fig:rhorN}
\end{center}
\end{figure}

The radial density profile holds for nearly an order of magnitude in particle number, producing marginal coring at inner radii, Fig.~\ref{fig:rhorN}. Low particle number does eventually produce a shift the outer power law and produces a complete coring of the inner halo at $N=1,000$ for both classical and Bose halos when using a softening length of $\epsilon=1$. Enlarging the softening length for the lowest-sampled simulations to $\epsilon=2$ puts the classical halo close to the universal profile, though the Bose halo requires $\epsilon =3$ to prevent the severe coring due mentioned above. The response of XC to softening will be a topic of future research.

\begin{figure}
\begin{center}
\includegraphics[width=\columnwidth]{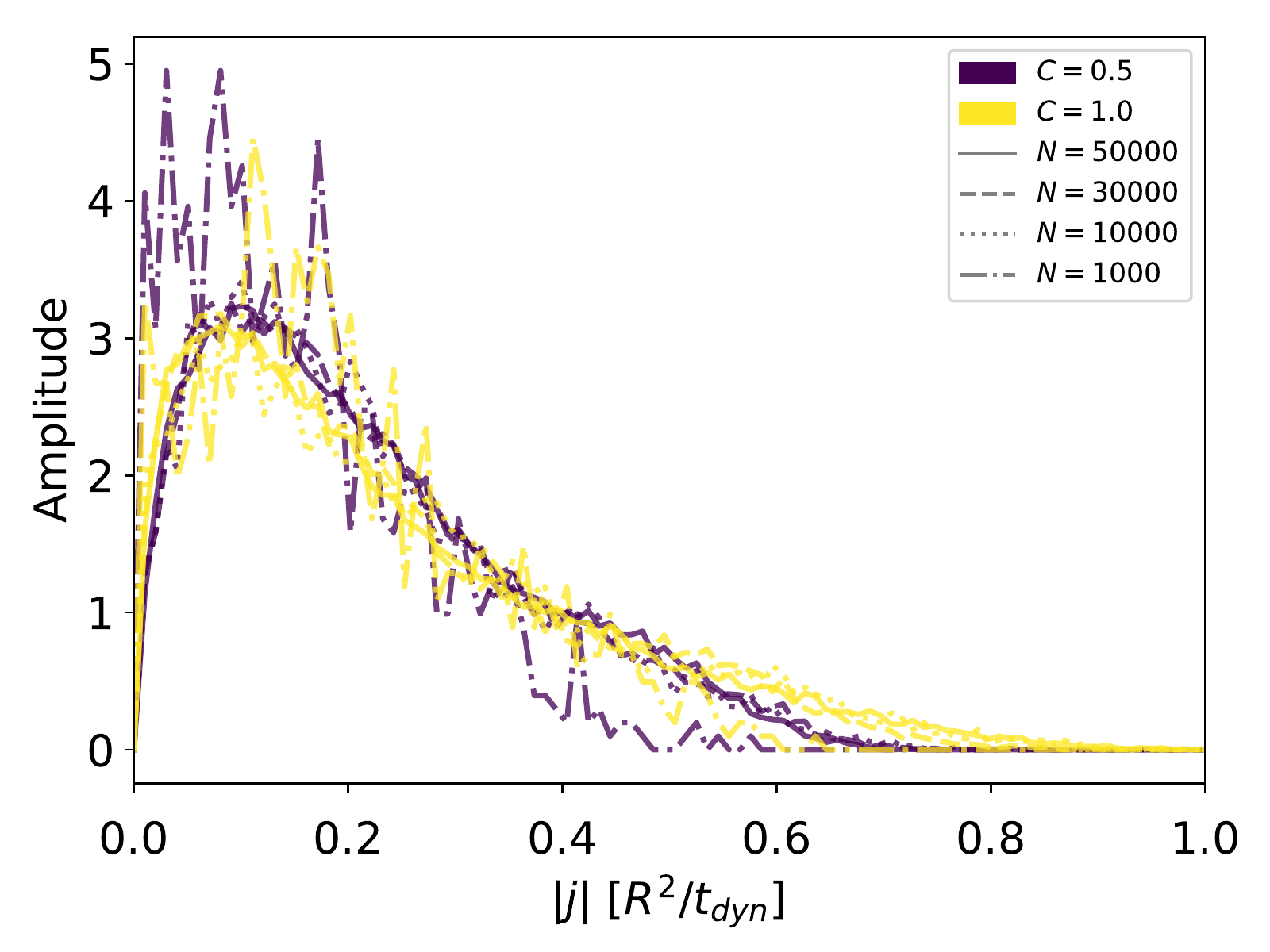}
\caption{Magnitude angular momentum distribution function of the Gaussian isolated collapse simulations of Sec.~\ref{SCS}. Halos were measure after evolving for $10 t_{dyn}$. Halos are without spin. Profile coloration indicates degree of correlation ranging from classical $C=1.0$ to highly correlated $C=0.5$. Line style indicates number of particles. Simulations above 10,000 samples use softening parameter $\epsilon=1$, while 1,000 particle simulations use $\epsilon=2$ for the classical case and $\epsilon=3$ for Bose.}
\label{fig:jdistrN}
\end{center}
\end{figure}

Distribution of the magnitude angular momentum are also consistent in the mean for halos of sampling above 10,000, Fig.~\ref{fig:jdistrN}. Binning noise over the mean signal is of Poisson type for these higher resolution halos, scaling with $1/\sqrt{N}$ samples per bin. Halos below the 10,000 particle mark are also seen to change width. Both $N=1,000$ simulations shown become narrower as force softening, which are set to $\epsilon=2$ for classical and $\epsilon=3$ for Bose, reduce the strength of the symmetry-breaking ROI which sets the spread in angular momentum. 

\begin{figure*}
\begin{center}
\includegraphics[width=\textwidth]{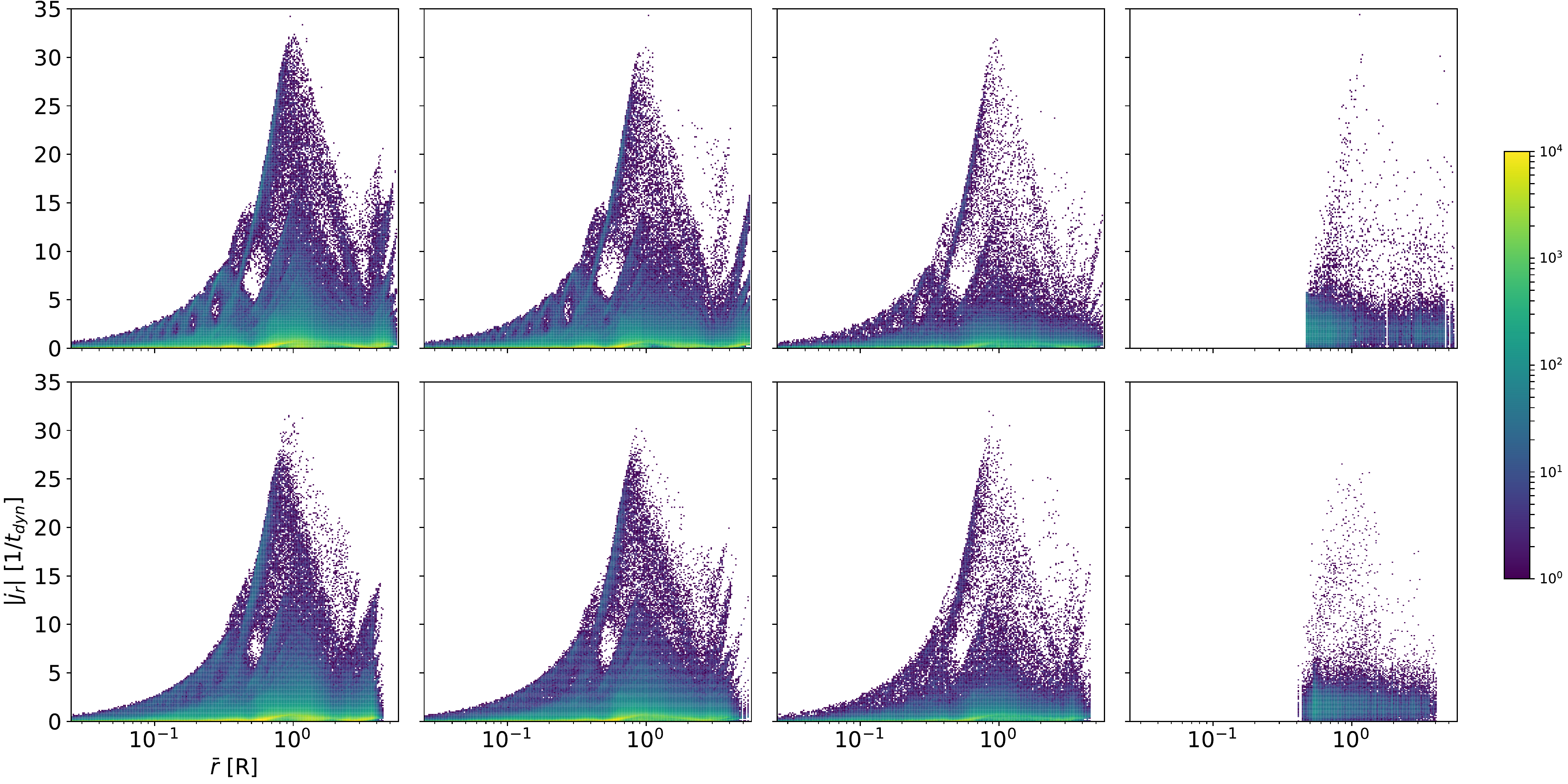}
\caption{Radial orbital action and mean orbit radius of Gaussian isolated collapse simulations. panels are organized by correlation (column-wise: 0.5,0.75,0.9,1.0 left to right) and spin (row-wise: 0.0,0.05,0.1 top to bottom). Samples are taken from $100$ equally-spaced frames during $10 - 12 t_{dyn}$. Sample points are taken such that each particle is given one mean radius and the radial auto-correlation power spectrum. Halos are without spin. Simulations above 10,000 samples use softening parameter $\epsilon=1$, while 1,000 particle simulations use $\epsilon=2$ for the classical case and $\epsilon=3$ for Bose.}
\label{fig:rmeanractionN}
\end{center}
\end{figure*}

Lastly, we observe the resonance structure observed in the radial orbital action, Fig.~\ref{fig:rmeanractionN}. Much of the structure discussed in the text at 50,000 samples remains visible at 30,000. Fine structure is more difficult to resolve for simulations with 10,000 samples. The smallest simulations have either no or too few particles with orbital radii much below $r=1R$ to see any virial resonances. Some of the resonant orbits not yet virialized can be made out even at $N=1,000$.

\bibliographystyle{mnras}
\bibliography{main.bib}

\end{document}